\documentclass[%
 aip,
 amsmath,amssymb,
 reprint,%
]{revtex4-1}

\usepackage{graphicx}
\usepackage{dcolumn}
\usepackage{bm}

\usepackage[utf8]{inputenc}
\usepackage[T1]{fontenc}
\usepackage{mathptmx}
\usepackage[vcentermath]{youngtab}
\usepackage{amsmath,amssymb,xspace}
\usepackage{xcolor}
\usepackage{physics,mathtools}
\usepackage{hyperref}

\begin{document}

\preprint{AIP/123-QED}

\title[Strongly interacting trapped 1D quantum gases: an exact solution]{Strongly interacting trapped one-dimensional quantum gases: an exact solution}

\author{A. Minguzzi}
\email{anna.minguzzi@lpmmc.cnrs.fr}
 \affiliation{Univ. Grenoble-Alpes, CNRS, LPMMC, 38000 Grenoble, France}
\author{P. Vignolo}%
 \email{patrizia.vignolo@inphyni.cnrs.fr}
\affiliation{ 
Univ. C\^ote d'Azur, CNRS, InPhyNi, 06560 Valbonne, France
}%

\date{\today}

\begin{abstract}
Quantum correlations can be used as a resource for quantum computing,
eg for quantum state manipulation, and for quantum sensing, eg for creating
non-classical states which allow to achieve the quantum advantage regime.
This review collects the predictions coming from a family of exact solutions
which allows to describe the many-body wavefunction of strongly correlated
quantum fluids confined by a tight waveguide and subjected to any form of
longitudinal confinement. It directly describes the experiments with trapped
ultracold atoms where the strongly correlated regime in one dimension has
been achieved. The exact solution applies to bosons, fermions and mixtures. It
allows to obtain experimental observables such as the density profiles and
momentum distribution at all momentum scales, beyond the Luttinger liquid
approach. It also predicts the exact quantum dynamics at all the times,
including the small oscillations regime yielding the collective modes of the
system and the large quench regime where the system parameters are changed
considerably. The solution can be extended to describe finite-temperature
conditions, spin and magnetization effects. The review  illustrates the
idea of the solution, presents the key theoretical achievements and the main
experiments on strongly correlated one-dimensional quantum gases.
\end{abstract}

\maketitle\




\section{Introduction}
Ultracold strongly correlated atomic gases are  extremely rich and complex physical systems. One needs to take into account the quantum degeneracy, the particle indistinguishability and their symmetry properties under exchange, the effects of the spin degrees of freedom and of the interactions.
The stronger is the interaction strength, the stronger are quantum correlations between the atoms and more difficult is to get an accurate description of the system, even numerically, and especially for the long-time dynamics. For these reasons, exact solutions for quantum systems are essential both for the deep understanding of fundamental physics and for the benchmark of classical and quantum simulators.

Exact solutions for one-dimensional (1D) homogeneous systems are well known in the literature.  Celebrated examples are the cases of 1D bosons or  fermions with contact interactions that are solvable by the Bethe Ansatz \cite{Lieb,Yang67,Sutherland68}. Such solution assumes crucially that the system is homogeneous, as described by a ring or a hard-wall trap. 

Several experiments on ultracold atoms, however, are performed in the presence of some type of external confinement, the most common being a  harmonic trap and/or optical lattices. For confined 1D systems, integrability generally breaks down. A remarkable  exception  is the infinitely repulsive limit where the absence of a length scale associated to interactions allows to obtain an exact solution for any form of external confinement. This is the case, for instance, of the Tonks-Girardeau gas (TG), a gas  of 1D bosons that can be mapped onto a system of spinless non-interacting fermions \cite{Gir1960}. Similarly, also multicomponent mixtures of bosons and fermions admit an exact solutions in the strongly repulsive limit. 

This review is dedicated to such a category of exact solutions, for trapped bosons, fermions and mixtures at zero and finite temperature. It complements the existing reviews on general features of one-dimensional systems of  ultracold atoms, specifically bosons \cite{Yurovsky2008,cazalilla2011onedimensional},  fermions \cite{guan2013fermi} and mixtures \cite{Sowinski2019}. 
We will discuss how to build exact solutions in the infinite repulsive limit and in its proximity, and we will compare exact results with mean-field approaches and virial expansion in the high-temperature limit.

\subsection{Experiments on 1D strongly correlated gases}
\begin{figure}[htb] 
\centerline{\includegraphics[width=0.65\linewidth]{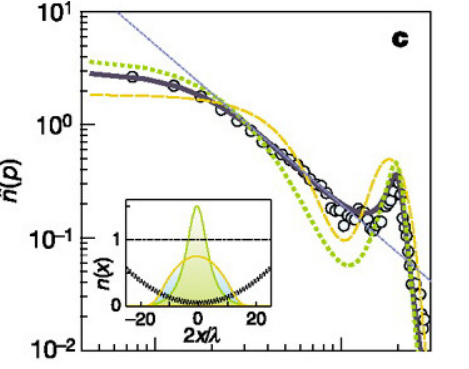}}
\caption[]{From [\onlinecite{Paredes04}]. Experimental observation of the TG regime for quasi-1D bosons in optical lattices. The measured momentum
  distribution is compared to the one predicted for a TG
  gas on a lattice (solid violet line), finding a good agreement. The
  predictions for the momentum distribution of an ideal Fermi gas
  (yellow dashed lines) and of an ideal Bose gas (green dotted line)
  are also shown for comparison. Reprinted by permission from Springer Nature Customer Service Centre GmbH: Paredes {\it et al.}, Nature {\bf 429}, 277 (2004), copyright 2004 (https://doi.org/10.1038/nature02530).}
\label{fig:tgexp2}
\end{figure}

\begin{figure}[tb]
\centerline{\includegraphics[width=0.5\linewidth,angle=270]{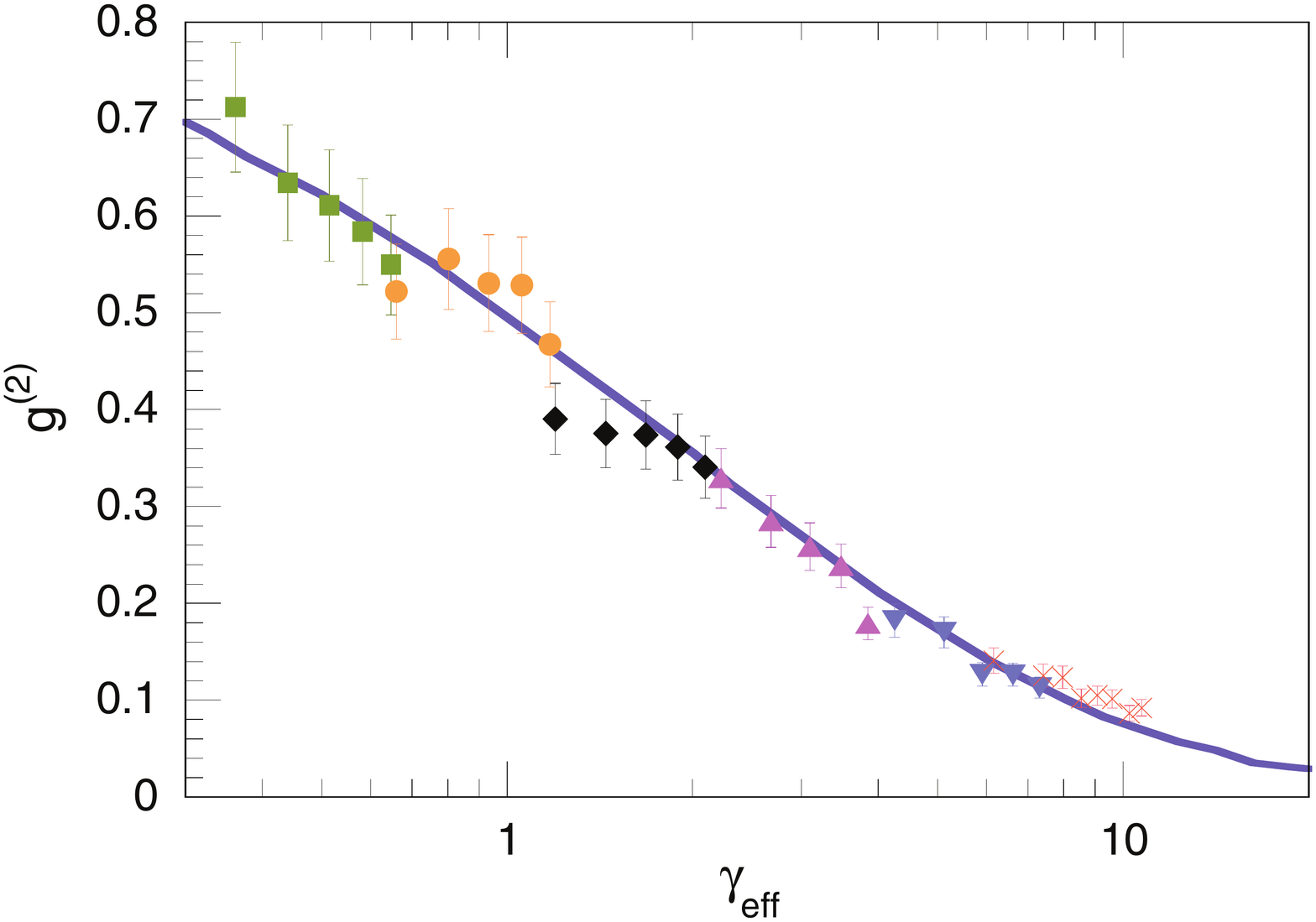}}
\caption[]{From [\onlinecite{Kinoshita05}]. Measurement of the local pair correlations  $g^{(2)}$  as a function of the coupling strength (data points) and 1D Bose gas theory from the Bethe Ansatz solution of the Lieb-Liniger model \cite{GanShlyap03} (solid line). The decrease of $g^{(2)}$ indicates approach to the strongly interacting regime; in the TG regime  $g^{(2)}$ is predicted to vanish as for noninteracting fermions.
Reprinted figure with permission from  [\onlinecite{Kinoshita05}],  \href{https://doi.org/10.1103/PhysRevLett.95.190406}{https://doi.org/10.1103/PhysRevLett.95.190406}. Copyright (2021) by the American Physical Society.}
\label{fig:tgexp3}
\end{figure}

The strategy to reach the strongly-correlated regime in 1D ultracold atomic gases consists in increasing the interaction strength with respect to the kinetic term: this can be  realized in several ways:  by means of Feshbach resonances \cite{Kinoshita04},  by increasing the atomic effective mass with the presence of a lattice potential \cite{Paredes04}, by decreasing the density, or by increasing the transverse confinement.
The first experiments achieving the strongly correlated regime in 1D confinements have been realized in the early 2000's both for bosons \cite{Kinoshita04,Paredes04} and  fermions \cite{Moritz2005}. 
The signature of the approaching TG regime for bosons was observed
in real space\cite{Kinoshita04} where fermionization shows up in the size cloud, and in momentum space \cite{Paredes04} where correlations manifest themselves (Fig.~\ref{fig:tgexp2}). In a subsequent experiment by Kinoshita et al.\cite{Kinoshita05}, the TG regime has been observed also in the strong decrease of two-body local correlations (see Fig.~\ref{fig:tgexp3}). Moreover, dynamical fermionization of bosons has been shown in the time evolution of the momentum distribution of an expanding cloud \cite{Wilson2020}. The control over experimental parameters is so accurate that it has been possible to make quantitative and stringent test of effective theories such as the Generalized Hydrodynamics \cite{malvania2021generalized}. 

\begin{figure}[htb] 
\centerline{\includegraphics[width=0.55\linewidth]{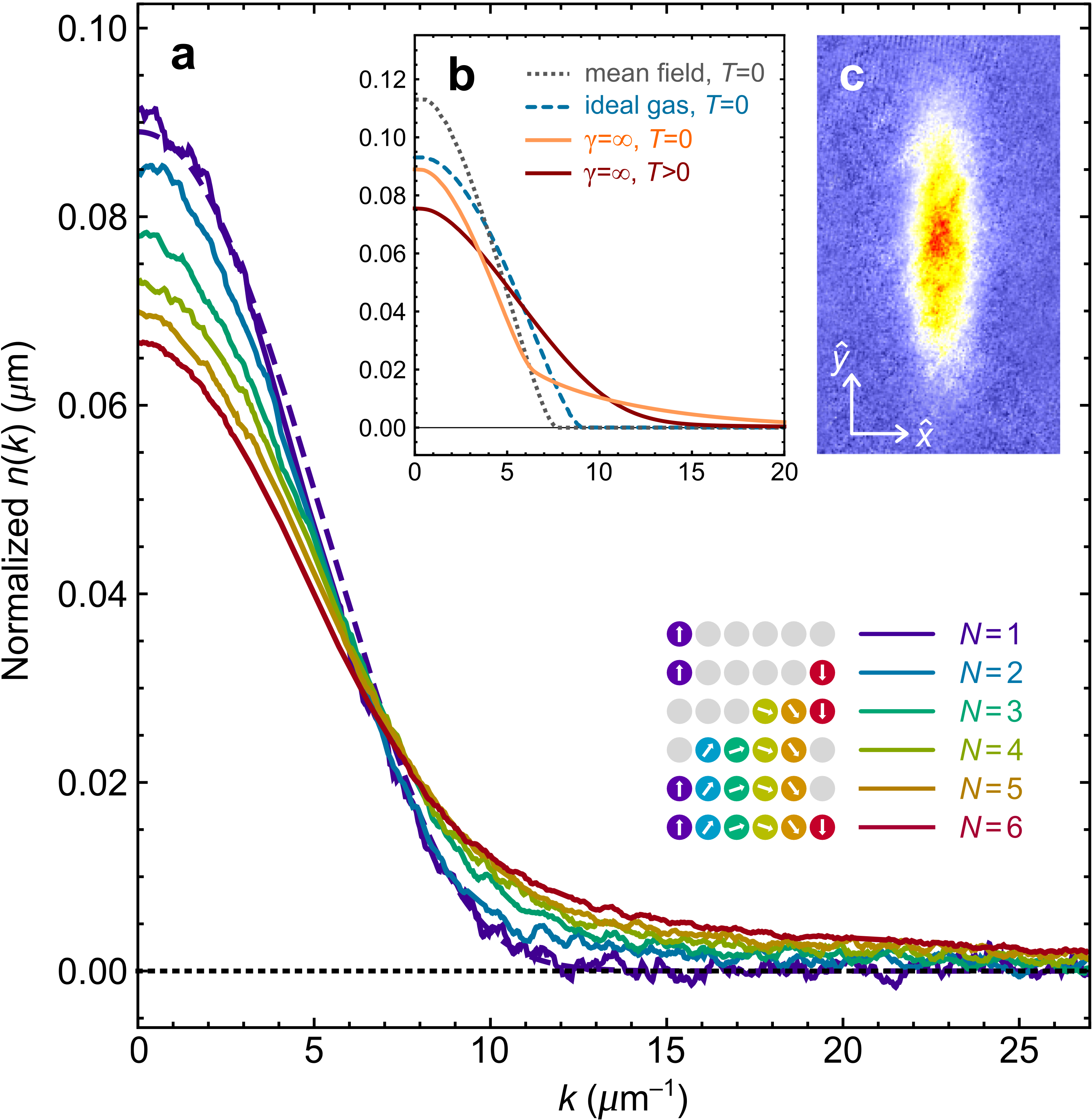}}
\caption[]{From [\onlinecite{Pagano2014}]. {(\textbf a)} Momentum distribution $n(k)$ measured with time-of-flight absorption imaging for different total number of atoms and the same atom number per spin component. {(\textbf b)} Theoretical $n(k)$ for the $N=2$ system derived from different models. {(\textbf c)} Averaged absorption image.
Reprinted by permission from Springer Nature Customer Service Centre GmbH: Pagano {\it et al.}, Nat. Phys. {\bf 10}, 198 (2014), copyright 2014 (https://doi.org/10.1038/nphys2878).} 
\label{fig:exp-pagano}
\end{figure}

Fermionization of fermions, namely the fact that strongly repulsive multi-component fermions behave like a non-interacting spinless fermionic gas, has been proven in Ref. [\onlinecite{Zurn2012}] for the case of a two-components mixture. Indeed the strongly correlated regime for ultracold gases can be obtained with multicomponent systems allowing {\it s}-wave scattering events at very low temperatures. Such mixtures offer the possibility to realize and study SU($\kappa$) systems (for instance $\kappa=6$ in Ref. [\onlinecite{Pagano2014}], see Fig.~\ref{fig:exp-pagano}) for balanced or imbalanced mixtures \cite{Liao2014}, paving the way for the study of  quantum magnetism and BCS-like pairing.  For instance, it has been proven that strongly-correlated fermions in a line are the experimental realization of a spin-chain Hamiltonian\cite{murmann2015antiferromagnetic}. The 1D Cooper pair mechanism has also been studied \cite{Zurn2013}.

Fluctuations depend on the interaction regime. Number fluctuations have been studied in the crossover from weak to strong interactions in Ref. [\onlinecite{Jacqmin2011}].
Typically, in ultracold atom experiments, the system is prepared in the ground state, but it has been shown that it is also possible to realize  a highly excited state with attractive interactions, the super-TG gas\cite{Haller2009}.



\begin{figure}[tb]
\centerline{\includegraphics[width=0.5\linewidth]{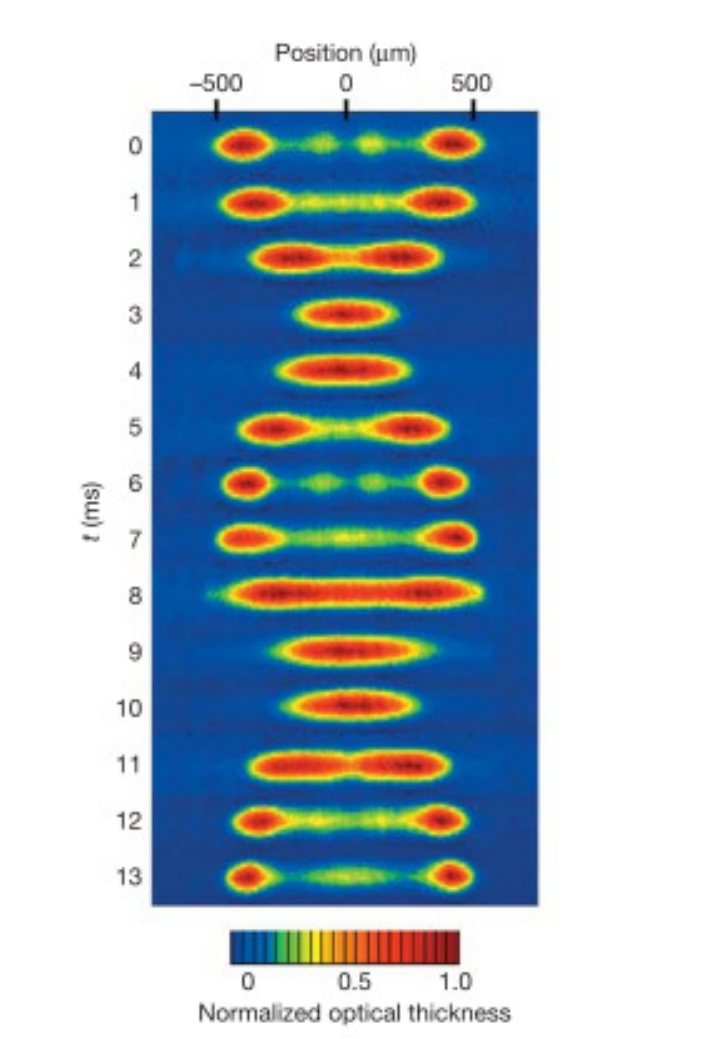}}
\caption[]{From [\onlinecite{Kinoshita06}]. Absence of damping in a quasi-onedimensional Bose gas driven out of equilibrium. Undamped oscillations are observed for an extremely long time, displaying the absence of collisional processes in quasi-1D gases. Reprinted by permission from Springer Nature Customer Service Centre GmbH: Kinoshita {\it et al.}, Nature {\bf 440}, 900 (2006), copyright 2006 (https://doi.org/10.1038/nature04693).}
\label{fig:tgexp4}
\end{figure}

Dynamical properties have then been explored in different setups. The dynamical structure factor has been studied in Ref. [\onlinecite{Fabbri2015}] and the non-equilibrium coherence dynamics in Ref.~[\onlinecite{Hofferberth2007}]. Ref.~[\onlinecite{Kinoshita06}] has shown
the absence of damping in a quasi-1D Bose gas driven out of equilibrium (see Fig.~\ref{fig:tgexp4}). This is the signature of the absence of collisions due to the reduced dimensionality: the oscillations have been shown to damp out if the dimensionality of the system is increased eg by allowing tunneling among the tubes of the 2D lattice.

Quantum dynamics of impurities in 1D Bose gases has been studied in Refs.[\onlinecite{Palzer09}],[\onlinecite{Catani2012}] and [\onlinecite{Meinert2017}].
An impurity,  dressed by the bosonic medium, feels an effective force which is smaller than the bare one\cite{Palzer09},
oscillates with an interacting-dependent amplitude\cite{Catani2012} and makes Bloch oscillations in the absence of a lattice, the ranging of the bosons playing the role of the lattice\cite{Meinert2017}.

The list of experiments we have presented in this section is not exhaustive, but shows the increasing interest of the community for 1D strongly correlated systems and the increasing control of the experimental techniques in order to realize and manipulate such systems.



\section{Methods}

\subsection{Exact solution for zero-temperature bosons}

\label{sec:exact-sol-bosons}

We consider $N$ bosons of mass $m$ in one dimension, interacting with contact interactions $v(x-x')=g \delta (x-x')$ and subjected to a longitudinal confinement $V_{ext}(x)$. 
The effective  interaction strength $g$ for the one-dimensional problem  can be expressed in terms of the three-dimensional interatomic $s$-wave scattering length $a_s$ and the typical scale for transverse confinement  $a_\perp=\sqrt{\hbar/m\omega_\perp}$ which is assumed to be harmonic with frequency $\omega_\perp$, leading to  \cite{Olsh98} $g=(4 \hbar^2 a_s/m a_\perp^2 )(1-\alpha a_s/a_{1D})^{-1}$ with $\alpha=1.4603...$. 
The Hamiltonian reads
\begin{equation}
H=\sum_{\ell=1}^{N}\left(-\dfrac{\hbar^2}{2m}\dfrac{\partial^2}{\partial x_\ell^2}+ V_{ext}(x_\ell)\right)+g\sum_{1\le \ell<j\le N}\delta(x_\ell-x_j).
\label{ham}
\end{equation}
This Hamiltonian corresponds to the Lieb-Liniger model \cite{LiebLiniger}. In absence of the external confinement, it can be solved by Bethe Ansatz for arbitrary interaction strength $g$. Here, we present an exact  solution holding for any external confinement due to Girardeau \cite{Gir1960}, valid  in the limit of infinitely  large repulsive interactions, ie for $g\rightarrow +\infty$, known as the Tonks-Girardeau (TG) limit.

The key idea of both the Bethe Ansatz and the Girardeau  solution is to replace interactions by a cusp condition on the many-body wavefunction $\Psi(x_1,... x_N)$,
\begin{equation}
\partial_x\Psi(x=0^+)-\partial_x\Psi(x=0^-)=\dfrac{mg}{\hbar^2}\Psi(x=0)
\label{cusp}
\end{equation}
where $x=x_j-x_\ell$ for any pairs of particles $\{j,\ell\}$.Then, in the limit $g=+\infty$ the many-body wavefunction must vanish at $x=0$. In this regime there is no energy or length scale associated to interactions and the many-body wavefunction can be built by mapping onto a non-interacting spinless Fermi gas wavefunction $\Psi_A(x_1,\dots,x_N)$
\begin{equation}
\Psi_A(x_1,\dots,x_N)=\dfrac{1}{\sqrt{N!}}{\det[\phi_{j}(x_\ell)]_{j,\ell=1,\dots,N}}
\label{slat}
\end{equation}
where the fermions are subjected to the same external confinement $V_{ext}(x)$, and the orbitals $\phi_{j}(x)$ are the solution of the single-particle Schr\"odinger equation ${\cal H}_0 \phi_j=\varepsilon_j \phi_j$ with eigenergies $\varepsilon_j$ and ${\cal H}_0 =-\dfrac{\hbar^2}{2m}\dfrac{\partial^2}{\partial x^2}+ V_{ext}(x)$.
The exact  many-body wavefunction for the TG gas then reads\cite{Gir1960}
\begin{equation}
\Psi(x_1,\dots,x_N)= \Pi_{j,\ell} {\rm sign} (x_j-x_\ell) \Psi_A(x_1,\dots,x_N)
\label{TGwf}
\end{equation}
where the mapping function ${\cal A}=\Pi_{j,\ell} {\rm sign} (x_j-x_\ell)$ ensures the bosonic symmetry under the exchange of two particles. In essence, interactions play the role of an effective Pauli principle and do not allow two particles to occupy the same spatial position or the same single-particle orbital. As we shall see below, there is a close connection among the Girardeau mapping and the Jordan-Wigner transformation \cite{jordan1928uber} introduced for lattice systems. 

The choice of the orbitals in the fermionic wavefunction allows hence to describe an arbitrary bosonic state. The ground state of the bosonic problem corresponds to a filled Fermi sphere of the mapped Fermi gas, with corresponding ground-state energy $E_{GS}=\sum_{j=1}^N \varepsilon_j$, with $j=1$ labelling the lowest-energy single-particle state.

The possibility of building a bosonic solution starting from a fermionic one is a signature of the statistical transmutation typical of one-dimensional systems. Extension to anyonic TG gases has also been studied \cite{Girardeau2006,Santachiara2008,Trombettoni2010,Patu2020}.  

In the uniform system of length $L$ with periodic boundary conditions corresponding to bosons on a ring, the ground-state many-body wavefunction reads\cite{Gir1960,Forrester03} 
\begin{equation}
\label{eq:psibox}
\Psi_{pbc}(x_1,...x_N)=(L^NN!)^{-1/2}\Pi_{1\le j<\ell\le N}2 i \sin(\pi (|x_j-x_\ell|)/L).
\end{equation}
Special care must be taken in this case depending on whether the number of bosons is even or odd: this is due to the fact that the mapping function ${\cal A}$ is periodic for odd $N$ and antiperiodic for even $N$: in the latter case one should use antiperiodic single-particle orbitals in order to ensure that the full manybody wavefunction is periodic.

Also in  the case of a harmonic confinement $V_{ext}(x)= m \omega_0^2 x^2 /2$ the wavefunction has an explicit solution. We have $\phi_{j}(x)=H_{j-1}(x/a_{ho})e^{-x^2/2 a_{ho}^2}/\sqrt{\pi 2^{j-1}(j-1)!}$, $\varepsilon_j=\hbar  \omega_0 (j-1/2)$ with $a_{ho}=\sqrt{\hbar/m\omega_0}$ the harmonic oscillator length and $H_j$ the Hermite polynomials.
By using the fact that for Vandermonde determinants $\det p_{j-1}(x_k)=\Pi_{1\le j<k\le N} (x_k-x_j)$ for $p_j(x)$ polynomial of degree $j$ with coefficient 1 in the $x^j$ term, and the properties of the Hermite polynomials, the wavefunction can be explicitly written as  \cite{GirWriTri01,Forrester03,Papenbrock2003}
\begin{equation} 
\label{eq:psitrap}
\Psi_{ho}(x_1,...x_N)=C_N\left[\Pi_{1\le j<\ell\le N}|x_j-x_\ell|\right]e^{-\sum_j (x_j/a_{ho})^2/2}
\end{equation}
where $C_N=\sqrt{N!} \left[\Pi_{m=0}^{N-1} 2^{-m} \sqrt{\pi} m!\right]^{1/2}$ is a normalization constant.

Quite remarkably, the above equation (\ref{eq:psitrap}) displays a striking connection to random matrix theory \cite{Dean2019}:  $|\Psi_{ho}(x_1,...x_N)|^2$ coincides with the joint distribution of the  eigenvalues of a $N\times N$ matrix belonging to the Gaussian unitary ensemble (GUE). This also implies that the distribution of the position of the last fermion in the trap is of Tracy-Widom (TW-GUE) type.

The Girardeau solution can also be extended to a class of time-dependent problems, where the particles are subjected to an arbitrary external time-dependent potential. Since the cusp condition must hold at all times, the solution by Girardeau can be extended to the time-dependent one \cite{GirWri00}
\begin{equation}
\Psi(x_1,\dots,x_N,t)= \Pi_{j,\ell} {\rm sign} (x_j-x_\ell) \Psi_A(x_1,\dots,x_N,t)
\label{TGwf-timedep}
\end{equation}
where the fermionic wavefunction is built with time dependent single-particle orbitals corresponding to the solution of the Schr\"odinger equation $ i \hbar \partial_t \phi_j=(-(\hbar^2/2m) + V_{ext}(x,t) \phi_j$ and $\phi_j(x,0)$ are the solution of the equilibrium Schr\"odinger equation in the initial potential $V_{ext}(x,0)$. This approach allows to study a large class of problems where the initial Fermi sphere evolves under the effect of a time variation of the external trapping potential. As an example, we may cite the expansion and interference  of a TG gas\cite{GirWri00}, the quench dynamics following a sudden change of the harmonic confinement \cite{Minguzzi05}  as well as periodically driven systems \cite{colcelli2019integrable}.

We also remark that the case of two hamonically trapped bosons has an exact solution for any value of the contact interaction strength $g$ \cite{Busch98}. In such a case,
$\Psi(x_1,x_2)=\phi_1(x_{CM})\psi_\nu(x_{rel})$, where $x_{CM}=(x_1+x_2)/2$, $x_{rel}=x_1-x_2$,
and $\nu$'s being the solutions of the transcendental equation
\begin{equation}
\dfrac{\Gamma(-\nu/2)}{\Gamma(-\nu/2+1/2)}=\sqrt{2}\dfrac{a_{1D}}{a_{ho}}.
\label{busch-nu}
\end{equation}
The wavefunction solving the Schr\"odinger equation for the relative motion~\cite{Busch98} reads
\begin{equation}
\psi_\nu(x_{rel})=
\dfrac{(\pi/2)^{1/4}}{\sqrt{a_{\textrm{ho}}}\sqrt{\mathcal{N}(\nu)}}\dfrac{2^{\nu/2}}{\Gamma(-\nu/2+1/2)} \, 
  \Phi\left(-\dfrac{\nu}{2},\dfrac{1}{2},\dfrac{x_{rel}^2}{a_{ho}^2}\right)
  \label{rizzi}
\end{equation}
where $\Gamma(u)$ is the gamma Euler function, $\Phi$ is the (Kummer) hypergeometric function, and
\begin{equation}
\mathcal{N}(\nu)= \Gamma (\nu +1)  \left\{ 1 + \tfrac{\sin (\pi  \nu)}{2\pi} 
	\left[ \mathtt{\Psi}\left(\tfrac{\nu}{2} + 1\right) - \mathtt{\Psi}\left(\tfrac{\nu}{2} + \tfrac{1}{2}\right) \right] 
	\right\}
\end{equation}%
is a normalization factor involving the digamma function $\mathtt{\Psi}(u) = \Gamma^\prime(u) / \Gamma(u)$.

In all the above examples, we have described a continuous system. The analogue of the Girardeau solution can also be formulated on the lattice in the limit of hard-core bosons.
In this case, specific commutation relations for the bosonic field operators $a_j$ have to be imposed to ensure the impenetrability condition, namely $[a_i,a_j^\dagger]=\delta_{i,j}$ for $i\neq j$ and $\{a_i,a_i^\dagger\}=1$, $a_i^2=0=(a_i^\dagger)^2$. 
In the presence of an additional harmonic potential $V_j=V_2 x_j^2$ on the lattice sites $x_j=j \Delta x$, with $\Delta x$ the lattice spacing,  the corresponding  Hamiltonian reads
\begin{equation}
H=-J \sum_j \left( a_j^\dagger a_{j+1} + H.c.\right) +V_2\sum_j x_j^2  a_j^\dagger a_{j},   
\end{equation}
where $J$ is the tunnel amplitude among nearest lattice sites. The exact solution then follows from the Jordan-Wigner transformation 
\begin{equation}
    f^\dagger_j=a^\dagger_j \Pi_{\ell=1}^{j-1}e^{-i \pi f^\dagger_\ell f_\ell} \,\,\, \,\,\,\,\, \,\,
    f_j= \Pi_{\ell=1}^{j-1}e^{i \pi f^\dagger_\ell f_\ell}a_j
\end{equation}
which maps the hard-core bosons onto non-interacting fermions described by the fermionic field operators $f_j$. Exploiting the Wick's theorem, this allows to obtain the one-body density matrix of hard-core bosons in a closed form requiring only the knowledge of the single-particle eigenstates of the Hamiltonian $H$ \cite{rigol2005groundstate} (see also [\onlinecite{Paredes2004}] for an alternative procedure). The above method has also been extended to obtain finite-temperature properties  \cite{rigol2005finitetemperature} and the spectral function \cite{settino2021exact} of lattice bosons.

\subsection{Specific focus on calculation of observables}
Given the knowledge of the many-body wavefunction, still it is not always immediate to obtain physical observables. We provide in this section specific examples on how to compute the main observables accessible in ultracold-atom experiments.

\subsubsection{Density profiles}
\label{Sec:dens-prof-method}
The first  observable accessible is the density profile 
that in terms of the bosonic creation and annihilation operators $\hat\Psi(x)$ and $\hat\Psi^\dagger(x)$
is defined as $n(x)=\langle \hat\Psi^\dagger(x)\hat\Psi(x)\rangle$.
For the TG gas, by using the Bose-Fermi mapping, the density profile,
as well as all other diagonal observables, is the same as the density profile
for a spinless fermionic gas.
If $\phi_j(x)$ are the single-particle orbitals for a non-interacting particles trapped
in an external potential $V_{ext}(x)$, then the density profiles can be written
\begin{equation}
n(x)=\sum_{j=1}^N|\phi_j(x)|^2=\sum_{j=1}^N\langle \phi_j|\delta(x-x_j)|\phi_j\rangle.
\label{eq:def-density-prof}
\end{equation}

Several elementary methods may be employed to obtain the density profile for a small particle number. Here we describe a Green's function method specifically taylored to address large systems. The key idea of the method is to express the $\delta$ function in Eq.~(\ref{eq:def-density-prof}) in terms of the imaginary part of the 
Green's function $G(x)=(x-\hat{x}+i\varepsilon)^{-1}$ for the position operator $\hat{x}$,
getting the expression
\begin{equation}
n(x)=-\dfrac{1}{\pi}\lim_{\varepsilon\rightarrow 0^+}{\rm Im}\sum_{j=1}^N\langle \phi_j|G(x)|\phi_j\rangle,
\label{green}
\end{equation}
and by using the relation between the trace of a matrix and the determinant of its inverse
\begin{equation}
n(x)=-\dfrac{1}{\pi}\lim_{\varepsilon\rightarrow 0^+}{\rm Im} \dfrac{\partial}{\partial\lambda}[\ln {\rm det }(x-\hat x+i\varepsilon+\lambda {\bf I})]_{\lambda=0}
\label{green2}
\end{equation}
where ${\bf I}$ is a diagonal matrix with the first $N$ diagonal elements equal to 1, and zero otherwise.
Expressions (\ref{green}) and (\ref{green2}) are particularly useful for the case of a harmonic trapping potential.
Indeed, for such a system, the position operator expressed on Hamiltonian eigenstates basis
is a tridiagonal matrix \cite{Vignolo2000}, whose non-zero elements takes the values
$[\hat x]_{j,j+1}=\sqrt{j/2}$. 
This implies that all techniques developed to deal with 1D tight-binding
systems for the calculation of the density of states can be exploited for the calculation
for the density profile of the harmonically trapped bosonic (fermionic) gas \cite{Vignolo2000}.
In particular one can use the Kirkman and Pendry relation \cite{Kirkman1984} in order to express
the density profile as a function of only the $G_{1,N}(x)$ Green's function element \cite{Vignolo2002}. This is somehow equivalent to the fact
that the total density can be expressed as a function of only
two wavefunctions, $\phi_N$ ane $\phi_{N-1}$ \cite{march2001}.
Let us underline that the Green's function  method sheds light on the quantum aspect of the system \cite{Vignolo2002}. Indeed one can observe that:
(i) if one reduces
the position operator to a $N\times N$ matrix, excluding all states that are not occupied 
by the fermions (at zero temperature), one obtains a density profile composed by $N$ delta peaks:
no exchange is possible among the particles; (ii) the Thomas-Fermi approximation,
that leads to a smooth density profile where shell effects are not visible, 
\begin{equation}
n_{\rm TF}(x)=\sqrt{2N-x^2/a_{ho}^2}/({\pi}a_{ho})
\label{TF-profile}
\end{equation}
corresponds
to a single occupied state (the first) 
and to fix the "hopping" terms $[\hat x]_{j,j+1}$ to the value $\sqrt{N/2}$ for any $j$.

The Green's function method can be generalized to higher dimensions \cite{Vignolo2003shell}. Alternatively it is also possible to express the exact density
profile in term of a sum over Laguerre polynomials \cite{Brack2001} or using random matrix theory \cite{Dean2016,Dean2019}. 
Three-dimensional shell effects have been numerically calculated in Refs.~[\onlinecite{Schneider1998mesoscopic}] and [\onlinecite{Bruun1998interacting}] and analytically
in any dimension \cite{mueller2004density} to leading order in $1/N$.

At arbitrary interactions, the harmonically trapped system is not integrable. Still, at sufficiently large $N$, the density profile can be obtained using the local-density approximation. For one-dimensional bosons at finite interactions, by using  the exact equation of state $\mu_{hom}[n]$ stemming from  the Bethe Ansatz solution for the homogeneous system\cite{Lieb}, one obtains the density profile using the implicit equation \cite{dunjko2001bosons,lang2017tan}
\begin{equation}
    \mu-V_{ext}(x)=\mu_{hom}[n(x)]
\end{equation}
together with imposing the normalization condition $\int dx \,n(x)=N$ which fixes the value of the chemical potential $\mu$ of the trapped system. As the Thomas-Fermi approximation in the TG limit, the LDA neglects the shell structure on the density profile.

\subsubsection{One-body density matrix}
The reduced one body-density matrix embeds the first-order coherence properties of the system. 
 It is defined as $\rho_1(x,y)=\langle \hat\Psi^\dagger(x) \hat\Psi(y)\rangle$.  The one-body density matrix  is a key quantity for  bosonic systems since its behaviour at large distances characterizes the off-diagonal long-range order, and in particular allows to obtain the condensate density $n_0$  according to the Penrose and Onsager criterion, which for  a homogeneous system reads  $\rho_1(x,y) \rightarrow n_0$ for $|x-y|\rightarrow \infty$.  In one dimensional systems there is no Bose-Einstein condensation, and the one-body density matrix decays as a power law for arbitrary (non-zero) interactions as predicted by the Luttinger liquid theory \cite{Haldane81} (see also \cite{Didier09b} for finite-size corrections). Finally, the one-body density matrix allows to obtain the momentum distribution of the gas by Fourier transform with respect to the relative coordinate,
 \begin{equation}
 n(k)=\int dx dy \, e^{i k (x-y)} \rho_1(x,y) .     
\label{eq:rho1-nk}
 \end{equation}

 For the TG gas a closed-form expression for the one-body density matrix was provided by Lenard \cite{Lenard}. Its asymptotic behaviour in the homogeneous system has been studied with several techniques:  at large distances as derived by the Lenard expansion \cite{vaidya1979oneparticle,vaidyaerratum,jimbo1980},  using replica trick \cite{gangardt2004universal} and random matrix theory \cite{Forrester03}, as well as the short distances  \cite{OlsDun03}, where a general connection has been found between the coefficients of the one-body density matrix and the local two- and three-body correlators \cite{olshanii2017connection}.

We detail here the calculation of the one-body density matrix for the TG gas, in a harmonic trap, following Ref.~[\onlinecite{Forrester03}]. In first quantization, the one-body density matrix reads
\begin{equation}
\rho_1(x,y)=N\int dx_2,... dx_N \Psi_{ho}^* (x, x_2,...x_N) \Psi_{ho}(y, x_2,....x_N).
\end{equation}
Inserting the expression for the TG many-body wavefunction (\ref{TGwf}), in analogy to [\onlinecite{Lenard}] it is possible to factor the many-body integrals into the determinant of one-body integrals, thus simplifying considerably the complexity of the calculation. 
The final result reads 
\begin{equation}
\rho_1(x,y)=c_N^2 e^{-(x^2+y^2)/2 a_{ho}^2} \det[ b_{j,k}(x/a_{ho},y/a_{ho})] 
\end{equation}
with 
\begin{eqnarray}
b_{j,k}&=&c_j c_k \int_{-\infty}^\infty dz e^{-z^2} (x-z)(y-z) z^{j+k-2}\nonumber \\
&-&2 \mathrm{sign}(x-y) \int_{x}^y dz e^{-z^2} (x-z)(y-z) z^{j+k-2}
\end{eqnarray}
and $c_j=[2^{j-1}/\sqrt{\pi} \Gamma(j)]^{1/2}$ .
The latter integral can be explicitly calculated in terms of the incomplete $\gamma$ function and  confluent hypergeometric functions \cite{Forrester03}. 
For large $N$, it can be shown that the one-body density matrix reduces to \cite{Forrester03}
\begin{equation}
    \rho_1(x,y)\sim N^{1/2} \frac{[G(3/2)]^4 }{\pi} \frac{(1-(x/R_{TF})^2)^{1/8}(1-(y/R_{TF})^2)^{1/8}}{|x/R_{TF}-y/R_{TF}|^{1/2}}
\end{equation}
where $R_{TF}=\sqrt{2N}a_{ho}$ is the Thomas-Fermi radius and $G(z)$ is the Barnes G-function. 
A useful extension of Lenard's method allows to obtain the expression of the time-dependent one-body density matrix,  $\rho_1(x,y,t)=\langle \Psi^\dagger(x,t) \Psi(y,t) \rangle$ with $\Psi(x,t) =e^{i H t}\Psi(x)e^{-i H t} $, whose dynamical evolution could be due e.g. to a quench of the system parameters. In such a case, using the time-dependent Bose-Fermi mapping and exploiting the properties of the determinants one has \cite{Pezer2007}
\begin{equation}
    \rho_1(x,y,t)=\sum_{j,\ell=1}^N \phi^*_j(x,t) A_{j\ell}(x,y,t) \phi_\ell(y,t),
\end{equation}
where the $N\times N$ matrix ${\mathbf A}(x,y,t)$ is given by ${\mathbf A}(x,y,t)=({\mathbf P}^{-1})^T \det \mathbf P $ and $P_{j\ell}(x,y,t)=\delta_{j\ell}-2 {\rm sgn}(x-y) \int_x^y dx' \,\phi^*_j(x',t) \phi_\ell(x',t)$.
This approach allows for a very efficient and  exact (up to numerical accuracy) calculation of the time evolution of the one-body density matrix.  
A finite-temperature extension of the above result has been demonstrated using Fredholm determinants approach \cite{atas2017exact}, valid each time the wavefunctions vanish at the boundaries of the system (eg for harmonically trapped or hard-wall potentials).

\subsubsection{Higher order correlators}
As we have seen in the previous section, the one-body density matrix provides
the momentum distribution that, together with the density profile, represents one of the main observable easily accessible in ultracold atoms.
Spatial and momentum fluctuations around the average are more difficult to be detected \cite{Clement2018},
but they can lead the most interesting signal as known in phase transitions or quantum optics.
Actually there is a particular interest in studying higher momentum occupation number, like the momentum distribution variance $\sigma_p^2=\langle N_p^2\rangle-\langle N_p\rangle^2$ and covariance $\langle N_pN_q\rangle$ \cite{Mathey2009noise,Kai2011scaling,Fang2016momentum,Bouchoule2012twobody,Lovas2017full,Lovas2017full,Devillard2020}, or the full counting statistics \cite{Lovas2017full,Devillard2020}.

The momentum distribution variance and coviarance are obtained by the Fourier transform 
of the two-body density matrix, that in first quantization reads
\begin{equation}
\begin{split}
&\rho_2(x_1,x_2;y_1,y_2)=\\
&\int dx_3\dots\int dx_N \Psi(x_1,x_2,x_3,\dots,x_N)\Psi^*(y_1,y_2,x_3,\dots,x_N).
\end{split}
\end{equation}
Indeed 
\begin{equation}
\begin{split}
\langle N_pN_q\rangle=&\int dx_1\int dx_2\int dy_1\int dy_2 \\&e^{ip(y_1-x_1)}e^{iq(y_2-x_2)}
\rho_2(x_1,x_2;y_1,y_2).
\end{split}
\end{equation}
For the homogeneous TG gas it has been found that, at any momentum $p$, with $p\ne0$ \cite{Lovas2017full,Devillard2020},
\begin{equation}
\langle N_p N_q\rangle=(1+\delta_{p,q})\langle N_p\rangle\langle N_q\rangle,
\end{equation}
and the full counting statistics of $N_p$ is exponential. The fact that there is no correlations
if $p\neq q$ is very different from the weakly interacting case where pairs with opposite momenta
are expected by the Bogolubov theory.
At $p=q=0$ the signature of quasi-long-range coherence is a decreasing function of fluctuations. In this case, it has been found that\cite{Devillard2020}
$\langle N_0^2\rangle=1.33\langle N_0\rangle^2$ 
and the full quantum statistics is neither exponential nor Gumbel as predicted in the weakly
interacting system \cite{Lovas2017full}, but seems to be in good agreement
with a positive-Gaussian distribution \cite{Devillard2020}.
In the trap system, it has been shown that non trivial correlations, including negative correlations appear for momenta smaller or of the order of the inverse radius of the gas \cite{Devillard2021}.

\subsubsection{Dynamical structure factor and spectral function}
The time-dependent Bose-Fermi mapping allows to obtain in an exact way dynamical correlation functions, as the dynamical structure factor  $S(k,\omega)$, and the spectral function $A(k,\omega)$. 

{\em Dynamical structure factor}. The dynamical structure factor is defined as the space-time Fourier transform of the connected density-density correlations
$S(k,\omega)= {\cal F} \langle n(x,t) n(y,0)\rangle_c$, with $\langle A B\rangle_c=\langle A B\rangle -\langle A \rangle \langle B \rangle$
 and $\langle ... \rangle$ denotes the quantum average on the state of the system.
It yields information on the spectrum of the system collective excitations, ie the response of the fluid upon the transfer of an energy $\hbar \omega$ and a momentum $\hbar k$.  This quantity is accessible in quantum gases experiments via Bragg scattering techniques \cite{Vignolo01,Golovach09,Clement2009}.

Since it involves only the density operators $n(x,t)$, the dynamical structure factor of the TG gas coincides with the one of a noninteracting Fermi gas. In the general case of a gas under external confinement, it reads
\begin{eqnarray}
S(k,\omega)=2 \pi \sum_{i,j}\left|\int dx \, e^{-i k x}\phi^*_i(x)\phi_j(x)\right|^2 \nonumber \\ \times f(\varepsilon_i) 
[1-f(\varepsilon_j)]\delta(\omega-\omega_{ij}) \;.
\label{s_i_kw}
\end{eqnarray}
with $\omega_{ij}=\varepsilon_j-\varepsilon_i$ and  $\phi_j(x)$ the single particle orbitals solution of the equilibrium Schr\"odinger problem. In the case of harmonic external potential we have \cite{Vignolo01}
\begin{equation}
{\cal S}(k, h)=2 \pi e^{-k^2/2}\sum_{i={\rm max}\{N-h,0\}}^{N-1}
\frac{i!}{(i+h)!} \left(\frac{k^2}{2}\right)^h 
\left[L_i^h\left(k^2/2 \right)\right]^2 \;.
\label{skw_exact}
\end{equation}
Here $h$ is an integer corresponding to a single-atom excitation
of $h$ quanta of the harmonic oscillator and  $L_i^h(x)$ is the $i^{th}$
generalized Laguerre
polynomial of parameter $h$.

It is useful to mention  that a local-density approximation for the dynamical structure factor yields a good approximation if $\omega \gg \omega_{0}$. The LDA reads $S_{LDA}(k,\omega)= \int dx\, n(x) S_{hom}(k,\omega;\mu[n(x)]$, with 
$\mu[n]$ being the equation of state, which in the TG gas reads $\mu[n]=\pi^2n^2/2m$, and $n(x)$ being the density profile of the gas.

{\it Spectral function}
The spectral function is defined as $ A(k,\omega) = - \frac{1}{\pi} \text{Im} G^R(k,\omega)$  where the retarded Green's function $G^R(k,\omega)$ is the Fourier transform of 
$G^R(x,t;y,t')=\theta(t-t')  \left[ G^>(x,t;y,t')-G^<(x,t;y,t') \right]$ with  $G^<(x,t;y,t')= -\imath \expval{  \hat \Psi^\dagger(y,t') \hat \Psi (x,t) }$,  and  $G^>(x,t;y,t')= -\imath\expval{ \hat \Psi (x,t) \hat \Psi^\dagger(y,t') }$
being the lesser and greater Green's functions. The spectral function contains the information on the response of the system when a particle is extracted or added to the system with wavevector $\hbar k$ and energy $\hbar \omega$. At difference from the dynamical structure factor, this quantity has bosonic character, ie it is very different from its fermionic counterpart. 

Thanks to the knowledge of the many-body wavefunction, it is possible to provide a closed, exact  expression for the spectral function of the TG gas in arbitrary external potential \cite{settino2021exact}. To give an idea of  the calculation, we provide some details  for the case of the lesser Green's function. We start from its definition,  
\begin{equation}
\begin{split}\label{eq:grLesI}
\imath G^< (x,t,y,t')_{\boldsymbol \eta}&= \expval{\hat \psi^\dagger (y,t') \hat \psi(x,t)}_{\boldsymbol \eta} \\
&=  \expval{ e^{i H t'} \hat \psi^\dagger (y) e^{-i H t'} e^{i H t} \hat \psi(x) e^{-i H t}}_{\boldsymbol \eta}
\end{split} 
\end{equation} 
where $\langle...\rangle_{\boldsymbol \eta}$ indicates the expectation value over the many-body state $|{\boldsymbol \eta}\rangle$  with ${\boldsymbol \eta}=\{\eta_1,...\eta_N \}$ single-particle quantum numbers, $H$ is the many-body Hamiltonian and $\hat \psi(x)$, $\hat \psi^\dagger(x) $ are bosonic field operators, satisfying the commutation relations $[\hat \psi(x),\hat \psi^\dagger(y)]=\delta(x-y)$.
Using the  completeness relation and setting  $X=x_2 \dots x_N$,  $Y=y_2 \dots y_N$, the  lesser Green's function in first quantization reads
\begin{equation}
\begin{split}
\imath G^< (x,t,y,t')_{\boldsymbol \eta} & = \frac{e^{i E_{\boldsymbol \eta}(t-t')}}{(N-1!)^2} \sum_n e^{-i E_{n}(t-t')}\int dY \Psi_{\boldsymbol \eta}^*(y,Y)\Psi_n(Y) \nonumber \\ &\times  \int dX \Psi_n^*(X) \Psi_{\boldsymbol \eta}(x,X).
\end{split} 
\end{equation} 
Using the exact expression for the many-body wavefunction of the state $\Psi_{\boldsymbol \eta}$ and for the excited states $\Psi_n$ belonging to the $N-1$ sector of the Hilbert space, we obtain the expression for the lesser Green's function of a TG gas: 
\begin{equation}\label{eq:lesserTG}
\imath G^< (x,t,y,t')=  \text{Det}[\textbf P (x,t)\textbf P (y,t')|_{{\boldsymbol \eta}{\boldsymbol \eta}}] a^<(x,t,y,t')
\end{equation}
where
\begin{align}
 a^<(x,t,y,t') &= {\boldsymbol \vec{\phi}(x,t)_{{\boldsymbol \eta}}^T} \  ( {[{\textbf P}(x,t) {\textbf P}(y,t')]^{-1}} )^T|_{{\boldsymbol \eta}{\boldsymbol \eta}} \   {\boldsymbol \vec{\phi}^*(y,t')}_{{\boldsymbol \eta}},
\end{align}
with $\text {P}_{lm}(x,t)= \int_{-\infty}^{\infty} \text{sign}(x-\bar x)\phi_l(\bar x,t)\phi^*_m(\bar x,t) d \bar x =\delta_{l,m} - 2 \ e^{-\imath t (\epsilon_l - \epsilon_m)} \int_{x}^{\infty} \phi_l(\bar x)\phi^*_m(\bar x) d \bar x$ and ${\boldsymbol \vec{\phi}}(x,t)= [\phi_1(x,t),\dots,\phi_M(x,t)]^T$.
As a comparison, the fermionic Green's function reads 
 $G^<_{F} (x,t,y,t') = \imath \sum_{\boldsymbol \eta} e^{ -\imath \epsilon_j (t-t')}\phi^*_j (y) \phi_j (x)$, ie it is diagonal in the particle space basis.
 We see then explicitely that the spectral function is a bosonic observable, ie does not coincide with the corresponding observable in the mapped Fermi gas.
 Notice that the above result recovers Pezer and Buljan result for the equal-time correlator \cite{Pezer2007}. Equation (\ref{eq:lesserTG}) hence allows for an efficient numerical implementation.
   A similar derivation can be done for $ G^> (x,t,y,t')$.  The above expression is fully general, ie it applies to any state $\eta$. Of special interest is the case when $\eta$ is the ground state, hence allowing to obtain the zero-temperature expression for $A(k,\omega)$.

\subsection{Multicomponent mixtures of bosons and fermions}
We consider a mixture of $N_B=N_1^B+\dots+N_b^B$ bosons and $N_F=N_1^F+\dots+N_f^F$ fermions,
with $N=N_B+N_F$, divided in $b$ and $f$ spin components \cite{Decamp2017}.
We assume a supersymmetric model where all particles have same mass $m$, all particles experience the same trapping potential $V_{ext}(x)$, and the interaction between particles $V_{int}(x,y)=g\delta(x-y)$ does not depend on the spin. This is compatible
with the fact that fermions with the same spin do not interact since the wavefunction being
antisymmetric for exchange of two identical fermions, the interaction term does not play any role.

The positions of the particles are given by the coordinates $x_1^{B,1},\dots,x_{N_B}^{B,b},x_{N_B+1}^{F,1},\dots x_N^{F,f}$. For the sake of simplicity of notations,
in the following, if not needed, we will omit the exponents specifying the type and the spin of the particles. Using this notation,
the Hamiltonian describing the system is formally the same as the one for the bosons (\ref{ham}).
As a consequence of the cusp condition (\ref{cusp}) that holds for the mixture too,
in the limit $g\rightarrow +\infty$ the many-body wavefunction vanishes whenever $x_i=x_j$. Thus it can be written \cite{Volosniev2014,Deuretzbacher2014}
\begin{equation}
\Psi(x_1,\dots,x_N)=\sum_{P\in S_N}a_P\theta_P(x_1,\dots,x_N)\Psi_A(x_1,\dots,x_N)
\label{vol}
\end{equation}
where $S_N$ is the permutation group of $N$ elements, $\theta_P(x_1,\dots,x_N)$ is
equal to 1 in the coordinate sector $x_{P(1)}<\dots<x_{P(N)}$, and $\Psi_A$
is the fully antisymmetric fermionic wavefunction (\ref{slat}).
The coefficients $a_P$ in Eq.~(\ref{vol}) for the case of the exchange of identical fermions are equal to 1,
and for the case of the exchange of identical bosons are equal to -1.
Thus the number of independent coefficients is reduced to
\begin{equation}
D_{N,b+f}=\dfrac{N!}{N_1!\dots N_{b+f}!}.
\end{equation}
We can thus reduce the dimensionality of our system from $N!$ to $D_{N,b+f}$ by regrouping
the sectors that are equal {\it modulo} permutations of identical particles. We call this basis
the snippet basis \cite{Deuretzbacher,Fang2011}.

In order to find all the other coefficients $a_P$ of the ground-state wavefunction we use a variational approach \cite{Volosniev2013}, by calculating the energy
at the first order with respect the small parameter $1/g$ (for a similar treatment on the lattice case see \cite{Ogata1990}) and by minimizing it in the limit
$g\rightarrow\infty$, 
\begin{equation}
E_g\simeq E_{g\rightarrow\infty}+\dfrac{1}{g}[\partial_{1/g}E]_{1/g\rightarrow\infty}=E_{g\rightarrow\infty}-\dfrac{1}{g}K.
\end{equation}
Remark that this corresponds to maximize $K$ that, as we will see in Sec. \ref{sec:mom} is proportional
to the Tan's contact \cite{Decamp2016-2}.
The procedure is the following. One write $K(a_P)$ by exploiting the cusp condition $(\ref{cusp})$,
\begin{equation}
K(a_P)=\sum_{P,Q\in S_N}(a_P-a_Q)^2\alpha_{P,Q},
\end{equation}
where
\begin{equation}
\label{Eq:alpha}
\begin{split}
\alpha_{P,Q}&=\int{\rm d}x_1,\dots {\rm d}x_N \theta_{\rm Id}(x_1,\dots,x_N)\delta(x_k-x_{k+1})
[\partial \Psi_A/\partial x_k]^2\\& \equiv \alpha_k 
\end{split}
\end{equation}
if $P$ and $Q$ are equal up to a transposition
of two consecutive distinguishable particles or indistinguishable bosons.
Then one imposes the stationarity of $K(a_P)$, taking into account the normalization condition $\sum_P a_P^2=1$,
by introducing the Lagrange multiplier $\lambda$
\begin{equation}
\partial_{a_P}[K(a_P)+\lambda\sum_P a_P^2]=0.
\end{equation}
This leads to the diagonalization problem
\begin{equation}
\tilde K {\bf a}_{\rm snip}=\lambda {\bf a}_{\rm snip},
\end{equation}
where ${\bf a}_{snip}$ is the vector of the $D_{N,b+f}$ independent $a_i$ coefficients,
and $\tilde K$ is a $D_{N,b+f}\times D_{N,b+f}$ matrix defined in the snippet basis by
\begin{equation}
\tilde K_{i,j}=\left\{
\begin{array}{ll}
-\alpha_{i,j} &{\rm if}\,i\ne j  \\
\sum_{d,k\ne i}\alpha_{i,k}+2\sum_{b,k\ne i}\alpha_{i,k} & {\rm if}\, i=j 
\end{array}
\right.
\end{equation}
where the index $d$ means that the sum has to be taken over snippets $k$ that transpose
distinguishable particles as compared to snippet $i$, while $b$ means that
the sum is taken over sectors that transpose identical bosons.
The $a_P$'s corresponding the highest eigenvalue of $\tilde K$ yield the ground state of
the system. The eigenstates corresponding to the other eigenvalues give access to the excites states
belonging to the same $D_{N,b+f}$ degenerate manifold at $g\rightarrow\infty$.

Our solution generalizes the one for the homogeneous system to arbitrary external potential in the hard-core limit $g\rightarrow \infty$. Indeed, in the homogeneous case (periodic boundary conditions or hard walls) the Bethe Ansatz provides an exact expression for the many-body wavefunction  of multicomponent supersymmetric mixtures at arbitrary interaction $g$ (see eg \cite{Guanrmp2013} for a review). Let us point out that the case of few trapped SU(2) fermions, with finite interactions, has been tackled in [\onlinecite{Gharashi2013}] and [\onlinecite{Lindgren2014}].

Finally, for fermionic SU$(\kappa)$ mixtures, with $\kappa$ the number of components, the above approach has been extended to time-dependent problems \cite{barfknecht2019dynamics}. In this case the solution is exact to order $1/g$, and is based on the mapping onto a Heisenberg SU$(\kappa)$ Hamiltonian \cite{Deuretzbacher2014,Yang2015,Yang2016} on a lattice of $N$ sites, where $N=N_1+...+N_\kappa$ is the total number of fermions:
\begin{equation}
H_s = (E_0 - \sum_i^{N-1} J_i)  + \sum_{i=1}^{N-1} J_i P_{i,i+1}. 
\label{eq:HamHeis}
\end{equation}
Here $P_{i,i+1}$ is the permutation operator among the particle $i$ and $i+1$, $E_0$ is the Fermi energy for the mixture, and  $J_i=\alpha_i/g$ is a site-dependent hopping amplitude related to the overlaps  $\alpha_i$ given in Eq.~(\ref{Eq:alpha}) above. An accurate expression for the exchange constants at large $N$ is given by \cite{Deuretzbacher2014} (see also [\onlinecite{Matveev2004}] and [\onlinecite{Matveev2008}])
\begin{equation}
J_i=\frac{\hbar^4 \pi^2 n_{TF}^3(x_i)}{3 m^2 g}
\end{equation}
where $n_{TF}$ is the Thomas-Fermi profile (\ref{TF-profile}) and $x_i$ is the center of mass of the $i-$th and $(i+1)-$th particle. 
The many-body wavefunction is then still described by Eq.~(\ref{vol}), but
with time-dependent coefficients $a_P(t)$ and $\alpha_i(t)$.
The time evolution of the coefficients $a_P$ is determined by the Heisenberg Hamiltonian (\ref{eq:HamHeis}). The time dependence of the coefficients $\alpha_i$ is due to the time variations of the external trapping potential  \cite{barfknecht2019dynamics}.
Purely spin dynamics, i.e. with $\alpha_i$ constant in time,  is realized by a suitable choice of the dynamical excitation protocol \cite{pecci2021universal}.

\subsubsection{Symmetry considerations}
Being the Hamiltonian (\ref{ham}) invariant with respect to permutation symmetry, its eigenstates
can be labelled via the Young diagrams corresponding to the irreducible representations of the permutation group $S_N$ \cite{Kerber}. As we shall see below, symmetry properties emerge in physical observables, e.g.~in the momentum distribution tails. 

In order to obtain the symmetry associated with a given wave function $\Psi(x_1,\dots,x_N)$
belonging to the degenerate manifold, we define a set of $D_{N,b+f}\times D_{N,b+f}$ matrices,
the $p$-cycle class-sums operators \cite{Kerber,Liebeck},
whose eigenvalues are directly connected to the irreducible representations of $S_N$, and thus to the Young tableaux. The $p$-cycle class-sum operator $\Gamma^{(p)}$ is the sum of the permutation of $p$ elements
in a cyclic way. For instance, the structure of $\Gamma^{(2)}=\sum_{i<j}(i,j)$ on the snippet
basis is the following:
\begin{itemize}
\item the diagonal elements $[\Gamma^{(2)}]_{A,A}$ are equal to $b_{N_b>1}-f_{N_f>1}$, $b_{N_b>1}$ and $f_{Nf>1}$ being the number of bosonic and fermionic components with a number of particles per component greater than one;
\item
the off-diagonal elements $[\Gamma^{(2)}]_{A,B}$ are equal to -1 (+1) if snippets $A$ and $B$ differ by the permutation of two distinguishable fermions (bosons);
\item the off-diagonal elements $[\Gamma^{(2)}]_{A,B}$ are equal to $-1^{\mathcal{N}_f}$ if snippets
$A$ and $B$ differ by the permutation of a boson with a fermion with in between $\mathcal{N}_f$ fermions;
\item the other off-diagonal elements are zero.
\end{itemize}
For the case of 4 particles $p_i$, with two particles per component $i$, $D_{N,b+f}=6$: $p_1p_1p_2p_2$, $p_1p_2p_1p_2$, $p_1p_2p_2p_1$, $p_2p_1p_1p_2$, $p_2p_1p_2p_1$, $p_2p_2p_1p_1$. If $f=2$ and $b=0$,
\begin{equation}
\Gamma^{(2)}_{N=4,f=2,b=0}=\left(\begin{array}{cccccc}
 -2& 1&-1&-1& 1&0  \\
 1&-2&1&1&0&1\\
 -1& 1&-2&0&1&-1\\
 -1& 1&0&-2&1&-1\\
 1&0& 1& 1&-2& 1\\
 0& 1&-1&-1&1&-2
\end{array}
\right)
\end{equation}
with eigenvalues $\gamma^{(2)}=\{-6,-2,-2,-2,0,0\}$.
These $\gamma^{(2)}$'s correspond to the Young tableaux
 (1,1,1,1)={\tiny\yng(1,1,1,1)}, (2,1,1)= {\tiny\yng(2,1,1)}, and (2,2)= {\tiny\yng(2,2)},
whose dimensions $d$ (1,3 and 2 respectively) is given by the Hook formula
\begin{equation}
d=\dfrac{N!}{\prod_{(i,j)}h(i,j)}
\end{equation}
where $h(i,j)$ is equal to the number of cells below the box $(i,j)+$ the number of cells at the right of the box $(i,j)$ +1.
Indeed the relation between the $\gamma^{(2)}$'s and a Young tableaux with a number of boxes $\lambda_i$ at line $i$ is
\begin{equation}
\gamma^{(2)}=\dfrac{1}{2}\sum_i[\lambda_i(\lambda_i-2i+1)].
\end{equation}

For the case of $f=0$ and $b=2$, i.e. a two-component Bose-Bose mixture, $\Gamma^{(2)}_{N=4,f=0,b=2}=-\Gamma^{(2)}_{N=4,f=2,b=0}$.
Its eigenvalues $\gamma^{(2)}=\{6,2,2,2,0,0\}$ correspond to
the Young tableaux (4)={\tiny\yng(4)}, (3,1)={\tiny\yng(3,1)}, and (2,2)= {\tiny\yng(2,2)},
again with dimensions 1,3 and 2.

For the case $f=1$ and $b=1$ \cite{Fang2011}, i.e. a Bose-Fermi mixture,
\begin{equation}
\Gamma^{(2)}_{N=4,f=1,b=1}=\left(\begin{array}{cccccc}
 0&-1&-1&1&1&0  \\
 -1&0&-1&-1&0&1\\
 -1&-1&0&0&-1&-1\\
 1&-1&0&0&-1&1\\
 1&0&-1&-1&0&-1\\
 0&1&-1&1&-1&0
\end{array}
\right).
\end{equation}
The eigenvalues $\gamma^{(2)}=\{-2,-2,-2,2,2,2\}$ correspond respectively to (2,1,1)= {\tiny\yng(2,1,1)}
and (3,1)={\tiny\yng(3,1)}, each one with dimension 3.
As we will see in the Sec. \ref{sec:res}, the ground state is always given by the most symmetric configuration (corresponding to the highest $\gamma^{(2)}$).

Remark that for $N>5$ the class-sum operator $\Gamma^{(2)}$ is not sufficient to label the Young tableaux and thus the symmetry of the system, since different diagrams can have the same eigenvalue $\gamma^{(2)}$.
In this case one needs to look at the other sum-class operators $\Gamma^{(p)}$, with $p>2$ in order
to uniquely identify the symmetry of a state. The general relation between the eigenvalues $\gamma^{(p)}$ and the Young tableaux with $n$ lines reads \cite{Katriel1993,Decamp2017}

\begin{equation}
\begin{split}
\gamma^{(p)}=&\dfrac{1}{p}\sum_{i=1}^{n}\mu_i(\mu_i-1)\dots(\mu_i-p+1)\\&\prod_{j\ne i}\dfrac{\mu_i-\mu_j-p}{\mu_i-\mu_j}
\end{split}
\end{equation}
where $\mu_i=\lambda_i-i+n$.

\subsection{Finite temperature bosons and fermions}

\subsubsection{Thermal Bose-Fermi mapping and diagonal observables}
The exact solution in the limit of infinite interactions both for TG bosons and multi-component fermions can be extended at finite temperature. 

We detail first the bosonic case. The key idea is that the Bose-Fermi mapping holds for any many-body energy eigenvalue, hence we can build a thermal density matrix for the TG gas in terms of the one of a Fermi gas. Writing the bosonic thermal density matrix as $\hat \rho_B= \sum_{N,\alpha} w_{N,\alpha} |\Psi_{B N \alpha} \rangle \langle\Psi_{B N \alpha} |$, with $w_{N,\alpha}=\exp[-\beta (E_{N,\alpha}-\mu N)]/Z$, $\alpha$ the state quantum number, $\beta=1/k_BT$, $\mu$ the chemical potential  and $Z=\sum_{N,\alpha} w_{N,\alpha}$, the expectation value of any observable $\hat O$ is given by $\langle \hat O \rangle= \sum_{N,\alpha} w_{N,\alpha} \langle\Psi_{B N \alpha} | \hat O  |\Psi_{B N \alpha} \rangle $. The  Bose-Fermi mapping for a given energy eigenstate $|\Psi_{B \alpha} \rangle$ states that it can be written in terms of the one of a noninteracting Fermi gas by the mapping operator $\hat A$, such that $|\Psi_{B N \alpha} \rangle= \hat A |\Psi_{F N \alpha} \rangle$.  In coordinate representation, for a $N$-particle state  $|\Psi_{B N \alpha} \rangle$  the many-body wavefunction reads
\begin{equation}
\label{Eq:psi_TG}
\Psi_{N\alpha}(x_1...x_N)=\Pi_{1\le j<\ell\le N} {\rm sign}(x_j-x_\ell) \Psi^F_{N \alpha}(x_1,x_2..,x_N).
\end{equation}
where $\alpha=\{\nu_1,...\nu_N\}$ is the set of single-particle quantum numbers, $  \Psi^F_{N \alpha}(x_1,x_2..,x_N) = \frac{1}{\sqrt{N!}} \det[\phi_{\nu_j}(x_k)]$  is the fermionic wavefunction constructed with the  single particle orbitals $\phi_{\nu_j}(x)$ corresponding single-particle energies $\varepsilon_{\nu_j}$, allowing to obtain the energy $E_{N \alpha}=\sum_{j=1}^N \varepsilon_{\nu_j} $.
The statistical Bose-Fermi mapping reads \cite{das2002interference}
\begin{equation}
\langle \hat O \rangle= \sum_\alpha w_\alpha \langle\Psi_{F \alpha} | A^{-1 }\hat O A |\Psi_{F \alpha} \rangle.
\end{equation}

Of specific interest are the observables that commute with the mapping operator $\hat A$, as the density and particle current operators. In this case their expectation value coincides with the fermionic one. For the particle density one has
\begin{equation}
n(x)=\frac 1 Z \sum_{N,\alpha} e^{-\beta (E_{N,\alpha}-\mu N)} n_{N\alpha}(x)
\end{equation}
with $ n_{N\alpha}(x)=\sum_{j=1}^N |\phi_{\nu_j}(x)|^2 $ which leads to 
\begin{equation}
n(x)=\sum_{j=1}^\infty f(\varepsilon_j) |\phi_j(x)|^2
\end{equation}
with $f(\varepsilon_j) =1/(e^{\beta (\varepsilon_j-\mu)}+1)$ the Fermi occupation numbers. The chemical potential at temperature $T$ and average particle number $N$ is obtained by setting $N=\sum_{j=1}^{\infty} f(\varepsilon_j)$.
An analytical expression for the thermal particle density at large $N$ is given by\cite{Dean2016}
\begin{eqnarray}
  n(x) \, &=& - \dfrac{1}{a_{ho}\sqrt{N}} \sqrt{\dfrac{k_B T}{ 2 \pi N \hbar \omega_0}} \nonumber \\
  &\textrm{Li}_{1/2}&\biggl(-(e^{ \frac{N \hbar \omega}{k_B T}}-1)\exp\left[-\frac{m\omega_0^2 x^2}{2k_B T}\right]\biggr).
  \label{eq_nT}
\end{eqnarray}
Similarly, the particle current density  of a TG gas at finite temperature coincides with the one of a Fermi gas and reads
\begin{equation}
j(x)=\frac{i \hbar}{2m}\sum_{j=1}^\infty f(\varepsilon_j) \left[\phi^*_j(x)\partial_x \phi_j(x) - (\partial_x \phi^*_j(x)) \phi_j(x)\right].
\end{equation}
Notice that all the above expressions can be used also to describe the time evolution of an initial thermal state, the time dependence being included in the evolution of the single-particle orbitals $\phi_j(x,t)$ \cite{das2002interference,cominotti2015dipole,polo2018damping,polo2019oscillations,dubessy2021universal}.

The same results hold for fermionized fermions and Bose-Fermi mixtures by a straightforward generalization of the previous mapping \cite{Capuzzi2020}. 

\subsubsection{Finite temperature one-body density matrix}
The Bose-Fermi mapping (\ref{Eq:psi_TG}) allows to construct the thermal average for the TG one-body density matrix :
\begin{eqnarray}
\label{Eq:rho1_initial}
\rho_{1B}(x,y)&=&\sum_{N,\alpha} P_{N,\alpha} N \int_I {\rm d}x_2,...{\rm d}x_N 
\nonumber \\
&\times&
\Psi_{N, \alpha}(x,x_2..,x_N)\Psi_{N, \alpha}^*(y,x_2,..,x_N).
\end{eqnarray}
Here  $I=(-\infty, \infty)$ is  the spatial integration domain, $P_{N,\alpha}=e^{-\beta(E_{N,\alpha}-\mu N)}/Z$ is the thermal distribution function, $Z=\sum_{N,\alpha}e^{-\beta(E_{N,\alpha}-\mu N)}$ the partition function for the TG gas with  $E_{N\alpha}=\sum_{j=1}^N \varepsilon_{\nu_j}$, $\beta=1/k_BT$, and $\mu$ the chemical potential.
Equation (\ref{Eq:rho1_initial}) can be simplified as illustrated in an
early work by Lenard  \cite{Len66}. The resulting expression reads \cite{vignolo2013}
\begin{equation}
\label{lenard_series_final}
\rho_{1B}(x,y)
=\sum_{j=0}^\infty \frac{(-2)^j}{j!} ({\rm sign}(x-y))^j 
\rho_{1B}^{(j)}(x,y)
\end{equation}
where we have defined
\begin{equation}
\rho_{1B}^{(j)}(x,y)=
\int_{x}^{y} {\rm d}x_2... {\rm d}x_{j+1} 
 \det [\rho_{1F}(x_i,x_\ell)]_{i,\ell=1,j+1},
\end{equation}
$\rho_{1F}(x,y)=\sum_{j=1}^N f_{j} \phi_{j}(x)\phi_{j}^*(y)$ being the {\em fermionic} one-body density matrix,  $f_j=1/[e^{\beta(\varepsilon_j-\mu)}+1]$  the Fermi occupation factor of a single-particle energy level, and 
in the above determinant one has to take $x_i=x$ for $i=1$ and $x_\ell=y$ for $\ell=1$. 
 Using the definition of the fermionic one-body density matrix,  the $j$-th term of the one-body density matrix is given by
\begin{eqnarray}
\rho_{1B}^{(j)}(x,y)&=&\!\!\!\!\sum_{\nu_1..\nu_{j+1}}  f_{\nu_1}... f_{\nu_{j+1}}\!\!\! \sum_{P\in {\cal S}_{j+1}} (-1)^P \phi_{\nu_1}(x) \phi_{\nu_{P(1)}}(y)\nonumber \\
&& \prod_{\ell=2}^{j+1}\int_y^x \!\! \!\! \!\!dx_\ell \, \phi_{\nu_\ell}(x_\ell)  \phi_{\nu_{P(\ell)}}^*(x_\ell).
\end{eqnarray}
This can be finally casted onto the compact form \cite{Len64,Pezer2007,Goold2008}
\begin{equation}
\label{Eq:rho1_operative}
\rho_{1B}^{(j)}(x,y)=\!\!\!\!\sum_{\nu_1..\nu_{j+1}} \!\!\!\! f_{\nu_1}... f_{\nu_{j+1}} \sum_{k=1}^{j+1}\phi_{\nu_1}(x) A_{\nu_1\nu_k}(x,y) \phi_{\nu_k}^*(y)
\end{equation}
where
$A_{\nu_1\nu_k}(x,y)$ can be expressed as functions of special functions for the case of a harmonically trapped gas \cite{vignolo2013}.

Another strategy to reduce the calculation of the thermal one-body density matrix to a simple double sum has been found by the authors of Ref. [\onlinecite{atas2017exact}].
Indeed it can be shown that one can write
\begin{equation}
\rho_{1B}(x,y)=\sum_{i,j=0}^{\infty}\sqrt{f_{i}}\phi_{i}(x)Q_{ij}\sqrt{f_{j}}\phi^*_{j}(y)
\label{bouch}
\end{equation}
where $Q_{ij}$ are the matrix elements of the operator 
\begin{equation}
Q(x,y)=(P^{-1})^T{\rm det}P
\end{equation}
with
\begin{equation}
P_{ij}(x,y)=\delta_{ij}-2{\rm sign}(y-x) \sqrt{f_{i}f_{j}}\int_x^y
dx''\,\phi_{i}(x'')\phi^*_{j}(x'').
\end{equation}
Using LDA combined with bosonization, it is possible to deduce an approximate expression for the thermal one-body density matrix at large $N$:
\begin{equation}
    \rho_{1B}(x,y)\simeq\frac{\mathcal{A}\, [n(x) n(y)]^{1/4}} {[\ell_T \sinh(\pi |x-y|/\ell_T)]^{1/2}},
  \label{eq_rho1_T-math}  
\end{equation}
where $\mathcal A$ is a constant, $n(x)$ is the thermal density profile given in Eq. (\ref{eq_nT}), and  $\ell_T=\hbar v_F/k_B T$ is the thermal length \cite{Cazalilla2004}, with $v_F=\hbar\pi n(0)/m$ being the Fermi velocity at the center of the trap.
It has been shown\cite{Devillard2021} that Eq. (\ref{eq_rho1_T-math}) gives results in very good agreement with those obtained by using the exact solution (\ref{bouch}) already for $N=10$.

\subsection{Momentum distribution and Tan's contact}
\label{sec:mom}

\subsubsection{Momentum distribution at small and large $k$}

The momentum distribution is one of the most common experimental observables, as it is measurable from time-of flight images of the atomic cloud after a sudden turn-off of the harmonic trap, assuming that interactions do not play any role during the expansion. This fact is ensured by the sudden drop of the density after switching off the confinement. 

From the theoretical point of view, the momentum distribution is readily obtained as the Fourier transform of the one-body density matrix (see again Eq.(\ref{eq:rho1-nk})).
The momentum distribution at small wavevector contains information about the off-diagonal long-range order: if there is a Bose-Einstein condensate, $n(k=0)=O(N)$ with $N$ the total particle number. In the case of a TG gas in a harmonic trap, it was shown that  \cite{Forrester03} $n(k=0)\propto \sqrt{N}$. The effect of interactions and quantum fluctuations is so strong to destroy Bose-Einstein condensation, and we have only quasi off-diagonal long-range order. For arbitrary interactions, it is possible to generalize the previous result\cite{Colcelli_2018} and one has $n(k=0) \propto N^\alpha$ with $\alpha\le 1/2$.
In the infinite size, homogeneous  system the  Luttinger liquid theory predicts the universal behaviour  $n(k)\sim k^{1/2K-1}$ for any interaction strength, where $K(\gamma)$ is the Luttinger parameter \cite{Cazalilla03} which depends on interaction strength. This results holds for wavevectors smaller than the inverse of the interparticle distance, ie $k\le n_0$ with $n_0$ the one-dimensional density. For finite systems of length $L$, the divergence is cut at small $k\sim 2\pi/L$.

At large wavevectors, another form of universality sets in and is related to contact interaction potential describing the ultracold gases. The one-body density matrix at short distance has a non-analytic behaviour at third order\cite{Forrester03}, i.e. $\rho_1(x,x')\sim |x-x'|^3$  implying a universal decay for wavevectors $k\ge n_0$ \cite{minguzzi02,OlsDun03} 
\begin{equation}
n(k)\sim k^{-4}.
\end{equation}
This property originates from the cusp at short distance in the many-body wavefunction due to the contact (delta)  interactions. It
holds in all spatial dimensions and both at zero and at finite temperature.


\subsubsection{Tan's contact}
The Tan's contact $\mathcal{C}$ is the weight of the large-momentum 
tails of the momentum distribution, 
\begin{equation}
\mathcal{C}=\lim_{k\rightarrow\infty} n(k)k^{4}.
\end{equation}
Tan's contact can be related to
 several many-body quantities, ranging from the interaction energy to the depletion rate by inelastic collisions, and many more \cite{Tan2008a,Tan2008b,Tan2008c,Zwerger2011}. 
Indeed it can be shown that
\begin{equation}
\mathcal{C}=\dfrac{gm^2}{\pi\hbar^4}\langle V_{int}\rangle=-\dfrac{m^2}{\pi\hbar^4}\dfrac{\partial E}{\partial g^{-1}},
\label{tan-tan}
\end{equation}
that in the $g\rightarrow \infty$ limit can be written 
\begin{equation} 
\mathcal{C}=\dfrac{m^2}{\pi\hbar^4}K.
\label{c-k-rel}
\end{equation}
We remind the reader that $K=-[\partial E/\partial g^{-1}]_{g\rightarrow\infty}$
and that we have shown that different values of $K$
label the multiplet of strongly interacting mixtures
(corresponding to the degenerate manifold in the limit $g\rightarrow\infty$)
and that the ground-state corresponds to the largest value of $K$.
This means that the different states with different symmetries can be labelled by the Tan's contact
and that the ground-state corresponds to the largest value of $\mathcal{C}$, namely to the
state with the largest momentum distribution tails.

\subsubsection{Tan's contact for a trapped TG gas: exact results}

In the TG limit, the contact does not depend
on the interactions, $K$ being the energy slope
in the $g \rightarrow \infty$ limit [Eq.~(\ref{c-k-rel}].
This allows to write $\mathcal{C}$, in this regime,
as a function
of the corresponding non-interacting fermionic two-body density matrix
$\rho_{2F}(x_1, x_2; x'_1, x'_2)$ \cite{Fang09}.
Indeed for a TG gas, it can be shown that
\begin{equation}
\mathcal{C}=\dfrac{2}{\pi}\int_{-\infty}^{+\infty} F(x)\,{\rm d}x
\label{bess}
\end{equation}
where we have defined
\begin{equation}
F(x)=\lim_{x',x"\rightarrow x}\dfrac{\rho_{2F}(x',x;x",x)}{|x-x'|x-x" |}.
\end{equation}
Eq.~(\ref{bess}) holds at zero \cite{Fang09} and finite temperature, both in the canonical \cite{Santana2019} and grand-canonical ensembles \cite{vignolo2013}.
$F(x)$ can be written explicitly as a function of the single-particle
orbitals $\phi_i(x)$,
\begin{equation}
F(x)=n(x)\sum_{i=1}^\infty|f_i\partial_x\phi_i(x)| ^2-|\sum_{i=1}^\infty f_i\phi_i(x)\partial_x\phi_i^*(x)| ^2.
\label{f-x}
\end{equation}
Eq.~(\ref{f-x}) is valid at zero temperature (where $f_i$ becomes a step function)
and at finite temperature in the grand-canonical ensemble \cite{vignolo2013}.
The explicit expression in the canonical ensemble has a similar but
more complicated structure \cite{Santana2019}.

Finally, remark that Eq.~(\ref{bess}) is the (non-homogeneous) TG limit
of the Lieb-Liniger contact expression
\begin{equation}
{\mathcal C}^{LL}=N(N -1) (m/\hbar^2)^2 g^2\rho_2(0, 0, 0, 0)
\label{gora}
\end{equation}
derived in [\onlinecite{GanShlyap03}]. Indeed since $\rho_2(0, 0, 0, 0)$
scales as $g^{-2}$, the TG limit of Eq.~(\ref{gora}) gives a finite value.

\subsubsection{Tan's contact for trapped mixtures at finite interactions} 

Let $\varepsilon$ be the ground-state energy density of a balanced homogeneous mixture and
$\gamma=mg_{1D}/\hbar^2\rho$ the dimensionless interaction parameter, $\rho$ being the total density. 
By using Eq.~(\ref{tan-tan}), one gets for the homogeneous gas
\begin{equation}
\mathcal{C}=g^2\dfrac{m^2}{2\pi\hbar^4}L\rho^2\dfrac{\partial e}{\partial\gamma}
\label{c-hom}
\end{equation}
where the
dimensionless average ground-state energy per particle $e$ is such that $\epsilon(\rho)=\frac{\hbar^2}{2m}\rho^3e(\gamma)$.

In the harmonically trapped system, it is possible to derive an expression for the contact
by performing a LDA \cite{OlsDun03}.
We define the energy functional $E[\rho]$
of the density $\rho(x)$ which, in the LDA, reads
\begin{equation}
E[\rho] = \int
dx [\epsilon(\rho) + (V_{ext}(x)-\mu)\rho(x)] . 
\label{tre}
\end{equation}
The ground-state
density profile is obtained by minimizing the energy functional, i.e. setting 
$\partial E/\partial \rho= 0$.
This yields an implicit equation for the density profile,
\begin{equation}
\dfrac{3}{2}\dfrac{\hbar^2}{m}\rho^2e(\gamma)-\dfrac{g\rho}{2}\dfrac{\partial e}{\partial \gamma}=
\mu-V_{ext}(x).
\label{quattro}
\end{equation}
The chemical potential $\mu$ is fixed by imposing the normalization condition $N =
\int dx \rho(x)$.
Combining Eqs. (\ref{tan-tan}), (\ref{tre}) and (\ref{quattro}),
we obtain Tan’s contact within the LDA:
\begin{equation}
\mathcal{C}_{LDA}=g^2\dfrac{m^2}{2\pi\hbar^4}\int dx\rho^2(x)\left.\dfrac{\partial e}{\partial \gamma}\right|_{\rho=\rho(x)}.
\label{eq-c-lda}
\end{equation}
Generally $e(\gamma)$ is not known analytically. However, for the Lieb-Liniger gas it exists
a very accurate analytical conjucture \cite{Lang2017} that allows to calculate the contact
at any interactions.
Moreover the asymptotic behaviour of $\lim_{\gamma\rightarrow\infty}e(\gamma)$
is known for the Lieb-Liniger gas as well as for balanced mixtures of SU($\kappa$) fermions.
The case of an imbalanced Fermi gas (a trapped 1D Fermi system
interacting with a single impurity) has been studied in [\onlinecite{Loft2016}].
In the strong-interacting limit, Eq.~(\ref{eq-c-lda}), takes the explicit form \cite{OlsDun03}
\begin{equation}
\mathcal C_{LDA,g\rightarrow\infty}^{LL} =\dfrac{\sqrt{2} \times128}{a_{ho}^3 45\pi^3}N^{5/2},
\end{equation}
for Lieb-Liniger bosons and
\begin{equation}
\mathcal{C}_{LDA,g\rightarrow\infty}^{SU(\kappa),F}=\mathcal{C}_{LDA,g\rightarrow\infty}^{LL}Z_1(\kappa)
\end{equation}
for SU($\kappa$) fermions \cite{Decamp2016-2}, where $Z_1(\kappa)=-\frac{1}{\kappa}[\psi(\frac{1}{\kappa})+C_{\rm Euler}]$,
$\psi$ being the digamma function and $C_{\rm Euler}$ the Euler constant.

In the limit of infinitely strong repulsions it is possible to obtain an exact formula for the contact
of a trapped mixture by using Eq.~(\ref{c-k-rel}). For each spin component $\sigma$, it reads\cite{Decamp2017}
\begin{equation}
    \mathcal{C}_\sigma(\infty)=\dfrac{m^2}{2\pi\hbar^4}
\sum_{\sigma'=1}^\kappa (1+\delta_{\sigma\sigma'})
\!\!\!\!\!\sum_{P\in \mathcal{G}_N(k||\sigma,\sigma')}\!\!\!\!\!\!\!(a_p-a_{P(k,k+1)})^2\alpha_k
\label{formula-c-jean}
\end{equation}
where $\mathcal{G}_N(k||\sigma,\sigma')$ is the subset
of permutations so that the indexes in positions $k$ and $k + 1$ correspond to particles belonging to components $\sigma$ and $\sigma'$.

At finite temperature, 
the contact can be derived by the thermodynamic form of the Tan's relation (\ref{tan-tan})
\begin{equation}
\left(\dfrac{d\Omega}{da_{1D}}\right)_{\mu,T}=\dfrac{\pi\hbar^2}{m}{\mathcal C},
\label{therm}
\end{equation}
where $\Omega$ is the grand thermodynamic potential.
For the homogeneous Lieb-Liniger gas $\Omega$ can be calculated by solving two coupled equations,
the Yang-Yang equations \cite{YangYang1969},
while for bosonic of fermionic mixture one needs in principle to deal with an infinity of coupled equations
that can be reduced to three complex coupled equations for the case of SU(2) fermions\cite{Patu2016} or bosons\cite{Patu2018}.
Once the grand thermodynamic potential is  known as a function of the density,
it is possible to calculate the contact for the thermal trapped system within the LDA scheme \cite{Yao2018,Capuzzi2020},
the principle of the calculation being exactly the same as for the zero temperature gas.
Fot the case of a thermal Lieb-Liniger gas and a SU(2) fermionic one, it has been shown that
the contact can be written as
\begin{equation}
\mathcal{C}_{LDA}=\dfrac{N^{5/2}}{a_{ho}^3} f(\xi_\gamma,\xi_T),
\label{LSP-work}
\end{equation}
where $f(\xi_\gamma,\xi_T)$ is a universal function depending only on the type of mixture and on the parameters $\xi_\gamma=a_{ho}/|a_{1D}|\sqrt{N}$ and $\xi_T=|a_{1D}|/\lambda_T$.
One can also write Eq.~(\ref{LSP-work}) under the form
$\mathcal{C}_{LDA}=N^{5/2}/a_{ho}^3 \zeta(\xi_\gamma,\tau)$ where $\zeta$ is a universal
function of $\xi_\gamma$ and $\tau=T/T_F$, or analogously as a function of two different combinations
of $\xi_\gamma$ and $\xi_T$.
This writing allows to deduce the scaling properties of the contact in the thermodynamic limit at finite temperature. We will discuss
this point deeply in Sec. \ref{two-is-enough}.

\subsubsection{Virial approach at large-temperature at strong interactions}
Let us start from Eq.~(\ref{therm}).
Using a virial expansion for $\Omega$,
 for a Lieb-Liniger gas one has
\begin{equation}
\mathcal{C}_{vir}^{LL}=\dfrac{4m\omega_0}{\hbar\lambda_T}N^2c_2
\label{virial-bos}
\end{equation} where $c_2=\lambda_T\partial b_2/\partial|a_{1D}|$,
and $b_2=\sum_\nu e^{ -\beta E_{rel,\nu}}$, $\lambda_T=\sqrt{2\pi\hbar^2/(mk_BT)}$ 
being the De Broglie wavelength.
For a harmonically-trapped gas 
$E_{rel,\nu}=\hbar\omega(\nu+1/2)$, where $\nu$'s are the solutions of the transcendental equation (\ref{busch-nu}).
In the TG regime, corresponding to $a_{1D}=0$, one has $\nu=2n+1$, with $n\in \mathcal{N}$.
In the strongly interacting regime $|a_{1D}|/a_{ho}\ll1$, we get the following
explicit expression for $\nu$ \cite{Yao2018}
\begin{equation}
\sqrt{\dfrac{2}{2n+1}}{\rm cot}(\pi\nu/2)\simeq\sqrt{2}\dfrac{|a_{1D}|}{a_{ho}}.
\end{equation}
This allows to obtain an analytical expression for $c_2$ as a function of $\xi_T$, 
\begin{equation}
c_2=\sqrt{2}\left(\dfrac{1}{2\pi\xi_T^2}-\dfrac{e^{1/2\pi\xi_T^2}}{2^{3/2}\pi\xi_T^3}{\rm Erfc}(1/\sqrt{2\pi}\xi_T)\right)
\label{mia}
\end{equation}
that is valid for {\it any} two interacting
particles with the same mass: two bosons, two fermions or one boson and one fermion.
Thus by inserting expression (\ref{mia}) in (\ref{virial-bos}), one gets for the
Lieb-Liniger gas
\begin{equation}
\mathcal{C}_{vir}^{LL}=\dfrac{2N^{5/2}}{\pi a_{ho}^3} \dfrac{\xi_\gamma}{\xi_T} \left(
\sqrt{2}-\dfrac{e^{1/2\pi\xi_T^2}}{\xi_T}{\rm Erfc}(1/\sqrt{2\pi}\xi_T)
\right),
\label{virial-mia}
\end{equation}
that in the TG limit simplifies to\cite{vignolo2013}
\begin{equation}
   \lim_{g\rightarrow\infty}\mathcal{C}_{vir}^{LL}= \dfrac{2\sqrt{2}}{a_{ho}^3}N^{5/2}\xi_\gamma\xi_T=
   \dfrac{N^{5/2}}{\pi^{3/2}a_{ho}^3}\sqrt{\tau}.
   \label{virial-TG}
\end{equation}
It is straightforward to show that \cite{Decamp2016-2,Capuzzi2020}, for the case of $N$ SU($\kappa$) balanced strongly interacting fermions
$\mathcal{C}_{vir}^{SU(\kappa),F}=\mathcal{C}_{vir}^{LL}(\kappa-1)/\kappa$.

\subsubsection{Scaling properties}
\label{two-is-enough}
Eq.~(\ref{LSP-work}) gives the scaling laws for the contact
at finite interaction and finite temperature in the thermodynamic limit ($N\gg 1$).
This means that we can calculate $f(\xi_\gamma,\xi_T)$
for $N=2$, for instance (if it easier to be calculated for $N=2$) 
and even if the result  once rescaled will not be applicable for small $N$, it will be valid for any $N\gg 1$. This because Eq.~(\ref{LSP-work}) has been derived
by the Bethe Ansatz equations and on the LDA on the top of the them, and both are valid at
large $N$. In order to cover the intermediate regime from few to many-body
one can use insted the scaling function
\begin{equation}
\tilde f(\xi_\gamma,\tau)=\dfrac{\mathcal{C}_N(\xi_\gamma,\tau)}{\mathcal{C}_N(\infty,\tau)}.   
\label{miamia}
\end{equation}
Equation (\ref{miamia}) 
holds in the regime of intermediate and large interaction ($\xi_\gamma\gtrsim 1)$
in the following situations: (i) at $T=0$ for a Lieb-Liniger gas and fermionic mixtures \cite{Rizzi2018}; (ii) for a Lieb-Liniger gas in the canonical ensemble at any temperature \cite{Santana2019}; and for a Lieb-Liniger gas and SU(2) fermionic mixtures in the grand-canonical ensemble at large temperature \cite{Santana2019, Capuzzi2020}.
This means that in these regimes the contact for $N$ particles in the $g \rightarrow \infty$ limit (the function $\mathcal{C}_N(\infty,\tau)$) contains all the $N$-dependency of the contact at almost any interactions.

\section{Results}
\label{sec:res}
\subsection{Density profiles}
\begin{figure} 
\begin{center}
\includegraphics[width=0.65\linewidth]{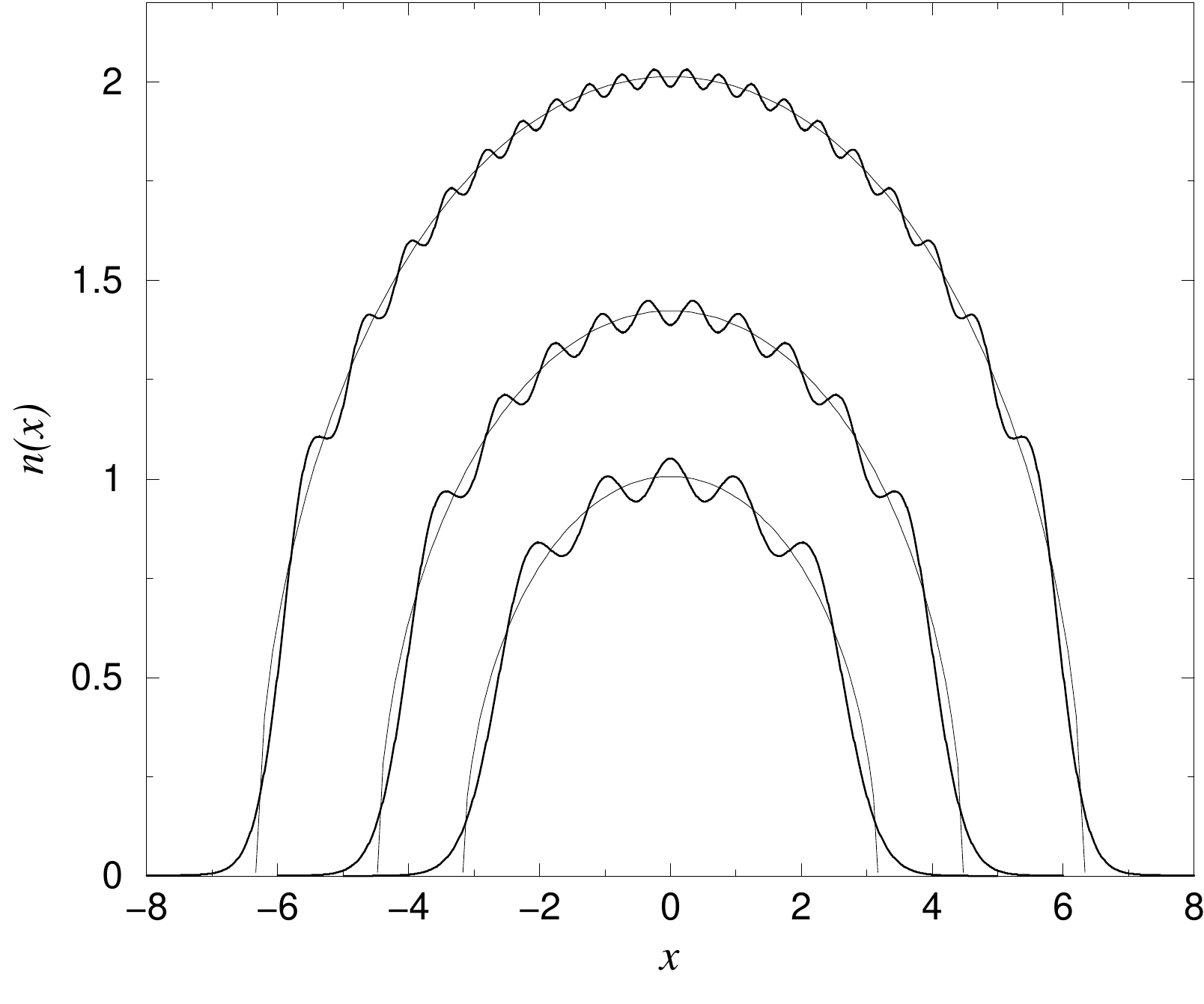}
\end{center}
\caption{\label{fig-F-2000} From  [\onlinecite{Vignolo00}].Exact particle density profile (bold lines) for $N=5$,
10 and 20 harmonically confined fermions, 
compared with the corresponding profiles
evaluated in the Thomas-Fermi approximation.
Positions are in units of the characteristic length
of the harmonic oscillator $a_{ho}=\sqrt{\hbar/(m \omega_0)}$ and
the particle density in units of $a_{ho}^{-1}$.
Reprinted figure with permission from  [\onlinecite{Vignolo00}],  \href{https://doi.org/10.1103/PhysRevLett.85.2850}{https://doi.org/10.1103/PhysRevLett.85.2850}. Copyright (2021) by the American Physical Society.}
\end{figure}

As outlined in Sec. \ref{Sec:dens-prof-method}, the density profile for $N$ TG bosons
is the same as the density profile for $N$ non-interacting fermions.
In Fig.~\ref{fig-F-2000} we show the results obtained with the Green's function method
for $N=5$, 10 and 20 fermions compared with the corresponding Thomas-Fermi density profiles [Eq.~(\ref{TF-profile})]. We observe a number of density oscillations equal to the number of particles,
whose amplitude decreases with $N$.
The position of these bumps correspond to the "classical" particles positions \cite{Vignolo2002},
namely to the position of the $N$ delta peaks that one would get cutting the $\hat x$ operator at
the first $N$ states, not considering highly unoccupied energy states.
This 
shell effect persists for 2D and 3D fermions trapped in highly anisotropic harmonic trap \cite{Vignolo2003shell} at zero temperature. Temperature washes out this effect already  at $k_BT\sim \hbar\omega_0$.

For the case of supersymmetric mixtures at infinitely strong interactions, the density profile can be 
calculated in an exact way by exploiting Eq.~(\ref{vol}). Let $x_i$ being the coordinate of a particle
of the component of the mixture we are interested in, then the density for that component
reads 
\begin{equation}
\begin{split}
n_\alpha(x)&=\sum_{j=1}^N \rho_\alpha^{(j)} \rho_j(x)
\label{vol2}
\end{split}
\end{equation}
with 
\begin{equation}
\begin{split}
\rho_j(x)&=\int dx_1\dots\int dx_N\,\delta(x-x_i)\\
&\times\theta_P(x_1,\dots,x_N)|\Psi_A(x_1,\dots,x_N)|^2,
\label{eq-rhoj}
\end{split}
\end{equation}
and $\rho_\alpha^{(j)}=\sum_{\{P\}}|a_P|^2$ being the $\alpha$-spin density probability at position $x_j$ \cite{deuretzbacher2016momentum}.
In the above expression $\{P\}$ are the elements of the snippet basis where a particle of spin $\alpha$ is in position $j$.
Remark that the $a_P$'s has the same symmetry of the trap. They are all equal in a box or in a ring trap,
while in the harmonic potential they satisfy the property $a_i=a_{N-i+1}$.
The resulting density profile depends on the state symmetry, but it can happen that
states corresponding to different symmetries have the same density profiles. This is the case, for instance, of TG bosons and non-interacting fermions. In the first case the many-body wavefunction is fully symmetric while in the second case it is fully anti-symmetric.

\begin{figure}
\begin{center}
\includegraphics[width=1\linewidth]{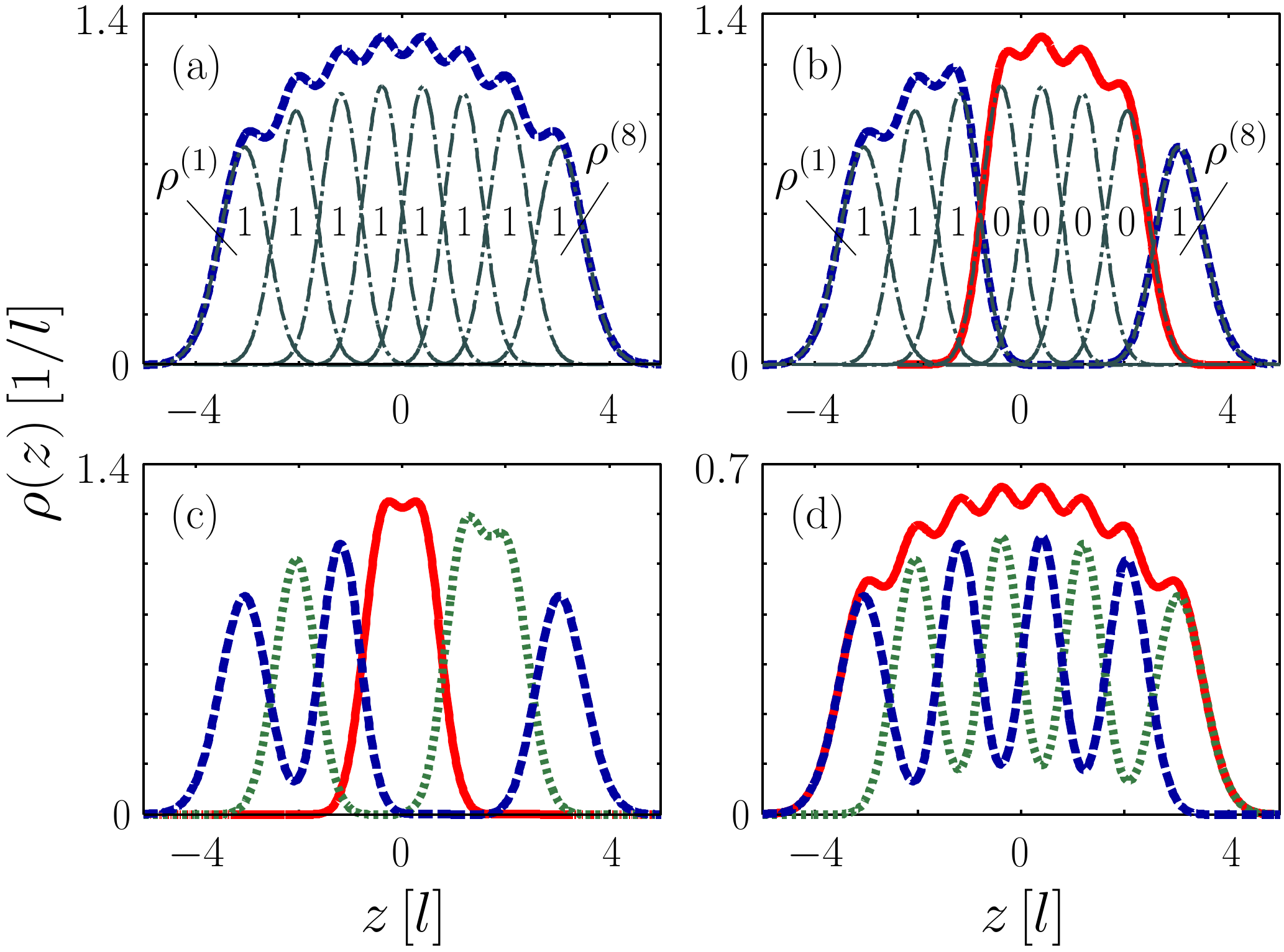}
\end{center}
\caption{\label{figBB-dens}From [\onlinecite{Deuretzbacher}].  Spin densities of 8 spin-1 bosons in different symmetry configurations. Shown are the densities $\rho_j$ (see Eq.~(\ref{eq-rhoj})
(gray dash-dotted line), and the components $n_0$ (red
solid line), $n_1$ (blue dashed line), and $n_{-1}$ (green dotted line) of
the spin density.
Reprinted figure with permission from  [\onlinecite{Deuretzbacher}],  \href{https://doi.org/10.1103/PhysRevLett.100.160405}{https://doi.org/10.1103/PhysRevLett.100.160405}. Copyright (2021) by the American Physical Society.}
\end{figure}

\begin{figure}
\begin{center}
\includegraphics[width=0.65\linewidth]{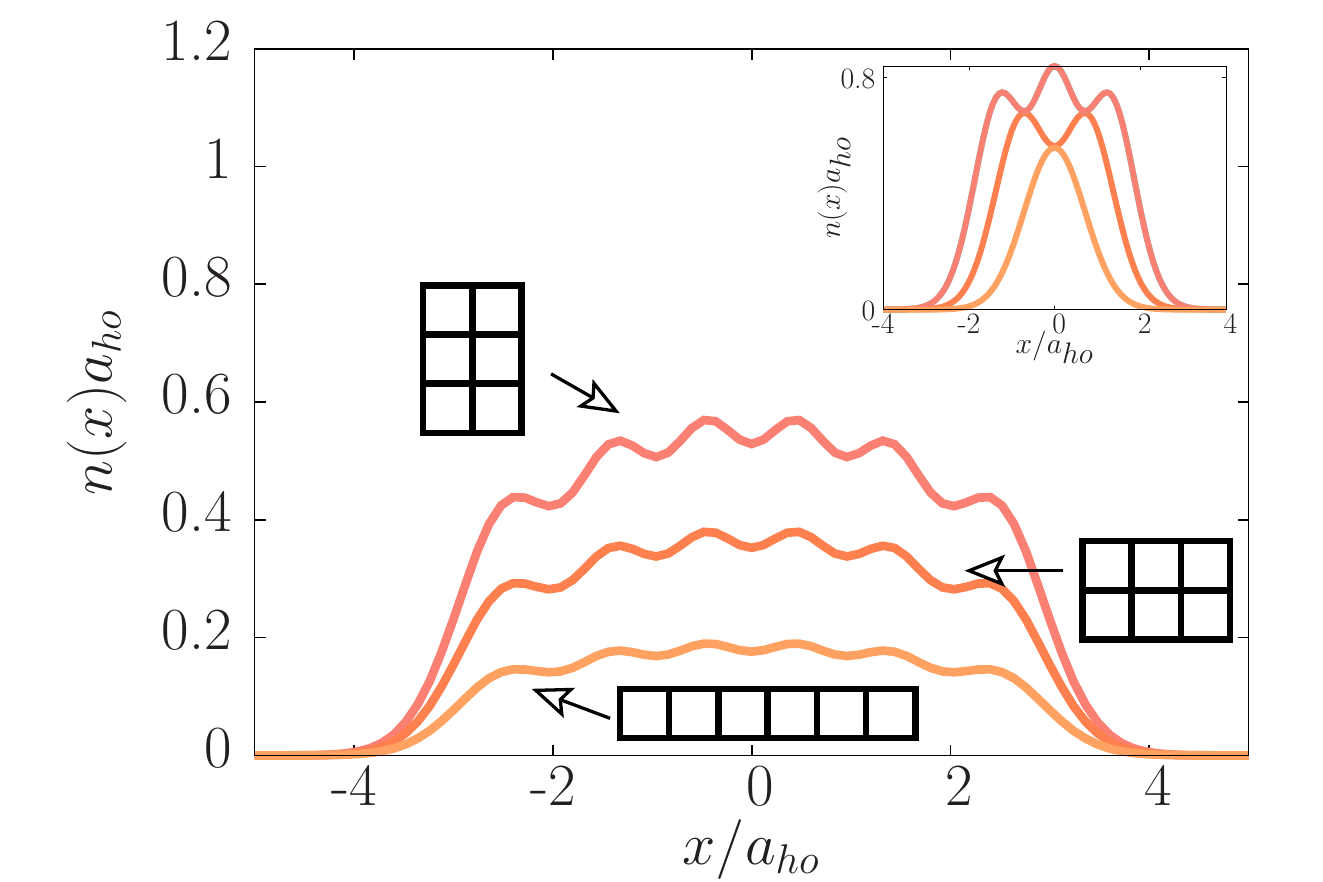}
\includegraphics[width=0.65\linewidth]{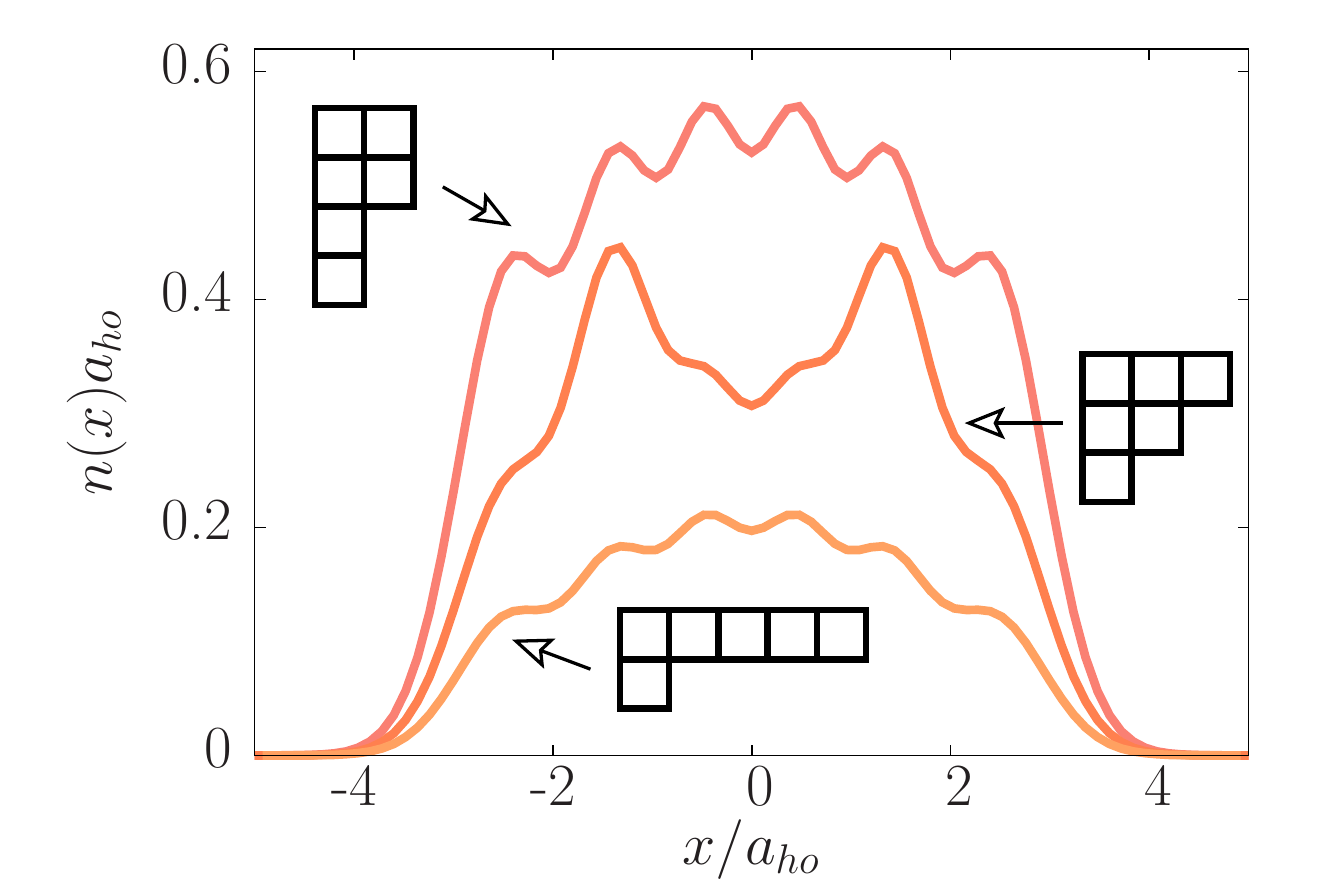}
\end{center}
\caption{\label{figFF-dens} From  [\onlinecite{Decamp2016}]. Density profiles for the ground state (top panel) and for the first many-body excited state with a symmetry different than the ground state (bottom panel) for three balanced mixtures (ie with the same number of particles in each species $N_1=\cdots=N_\kappa$) of strongly interacting Fermi gases, with different numbers of components $\kappa=2,3,6$ and total particle number 
 $N=6$ (from top to bottom: $N_\nu=3,2,1$).  The density profiles are the same for each component of the mixture. The inset shows the corresponding ground state density profiles for the case of the corresponding mixtures of noninteracting fermions. Reprinted with permission from Decamp {\it et al.}, New J. Phys. {\bf 18}, 055011  (2016). Copyright 2016, (https://doi.org/10.1088/1367-2630/18/5/055011). Author(s) licensed under a Creative Commons Attribution 4.0 License.}
\end{figure}

\begin{figure}
\begin{center}
\includegraphics[width=1\linewidth]{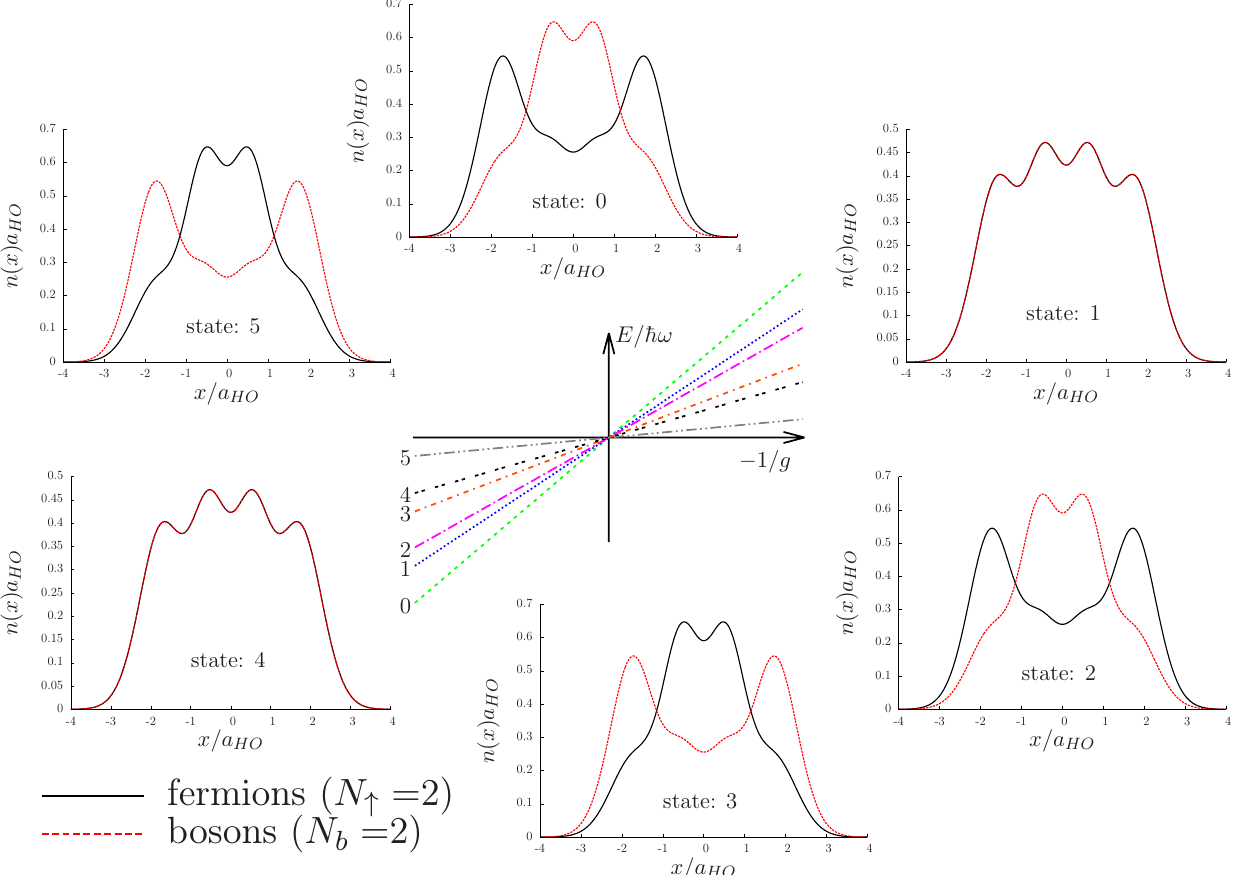} 
\end{center}
\caption{\label{figBF-dens}From [\onlinecite{Dehkharghani_2017}].  Density profiles for the six states with different symmetry for a mixture of 2 fermions and 2 bosons. Republished with permission of IOP Publishing, from {\it Analytical and numerical studies of {B}ose{\textendash}{F}ermi mixtures in a one-dimensional harmonic trap}, A.S. Dehkharghani {\it et al.}, Journal of Physics B: Atomic, Molecular and Optical Physics {\bf 14}, 144002 (2017); permission conveyed through Copyright Clearance Center, Inc.}
\end{figure}

In Figs.~\ref{figBB-dens}, \ref{figFF-dens} and \ref{figBF-dens} we show the density profiles for several harmonically trapped mixtures.
Figure \ref{figBB-dens} shows the shape of the $\rho_j(x)$'s [Eq.~(\ref{eq-rhoj})] for the case of 8 trapped particles, and the spin-density profiles for a spin-1 boson gas, for different symmetry configurations. 

Figure \ref{figFF-dens} focuses on balanced fermionic mixtures with $N=6$ and $\kappa=2$, 3 and 6 number
of components. The symmetry of the many-body state is indicating by the corresponding Young tableau.
The density for each component of the ground state (top panel of Fig.~\ref{figFF-dens}) is the same for any mixture up to a normalization factor. This is somehow due to the fact that their symmetry corresponds to a Young tableau (in the tree cases) that is a closed box.
As soon as this sort of "Young tableau symmetry" is broken, the density profiles may change deeply for each component as shown for the case of excited states in the bottom panel of Fig.~\ref{figFF-dens}. 
The density profiles for boson-fermion mixtures have been studied in Refs.[\onlinecite{Hu2016}], [\onlinecite{Zinner2017}] and [\onlinecite{Decamp2017}]. Some examples are shown in Fig.~\ref{figBF-dens}. In the case of a boson-fermion mixture
the ground state (the state corresponding to the largest slope energy) can never correspond to a "symmetric" Young tableau and the different components
in the same mixture have different density profiles. For such a state (state 0, top panel), one observes a demixing behaviour with
the bosonic components concentrates in the center of the trap and the fermionic ones more occupying the peripheral region of the trap.

\subsection{Dynamical structure factor and spectral function}

The TG solution allows also to access in an exact way to dynamical properties. In this section we focus on linear response regime, eg the response of the fluid to small perturbations. Large quenches and strongly out-of-equilibrium dynamics will be treated in Sec.~\ref{Sec:quenches} below. 

For a strongly correlated gas, both the spectral function and the dynamical structure factor are 
 non-vanishing in a large part of the $(k,\omega)$ plane. Since the particle are interacting and fill an effective Fermi sphere, there are several ways for to adjust an excitation with transfer of a given momentum $\hbar k$ and energy $\hbar \omega$, noticeably by single or multiple particle-hole excitations. Due to the underlying Fermi sphere structure induced by correlations, in the homogeneous system there are also regions of the $(k,\omega)$ plane which are kinematically forbidden \cite{ImaDemAnn}, for example, in the case of dynamical structure factor,  at finite momentum $0<k<2k_F$ and small frequency, the first excitations possible are those who correspond to the backscattering processes $-k_F$ to $k_F$ around the Fermi points.

In the homogeneous system, the non-linear Luttinger liquid approach \cite{Imambekov1,ImaDemAnn} provides a complete description of the spectral properties in proximity of each singularity line. Complementary to that approach, we provide here an exact analysis of the trapped case.

\begin{figure}
    \centering
    \includegraphics[width=0.45\linewidth]{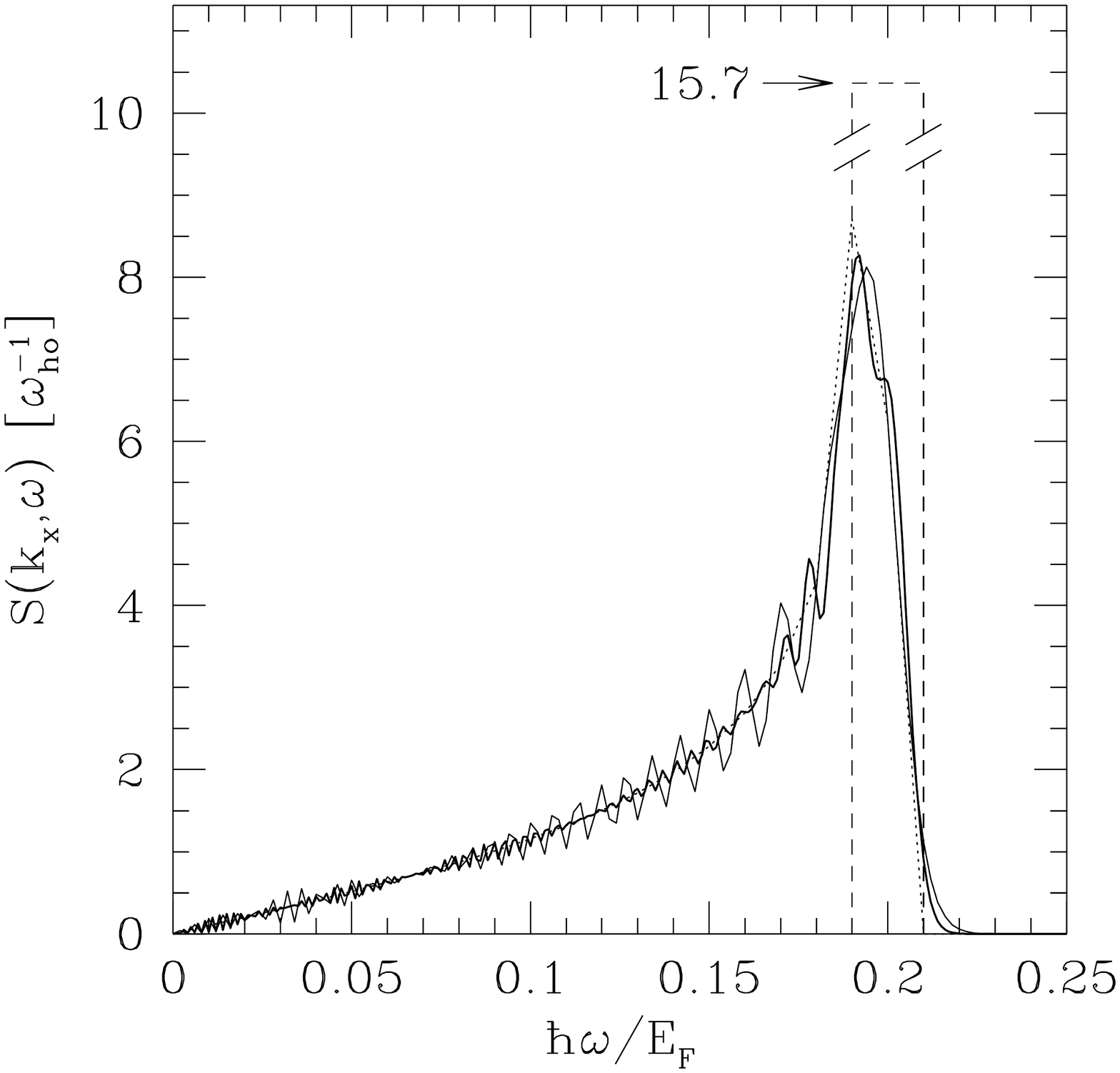}
    \includegraphics[width=0.45\linewidth]{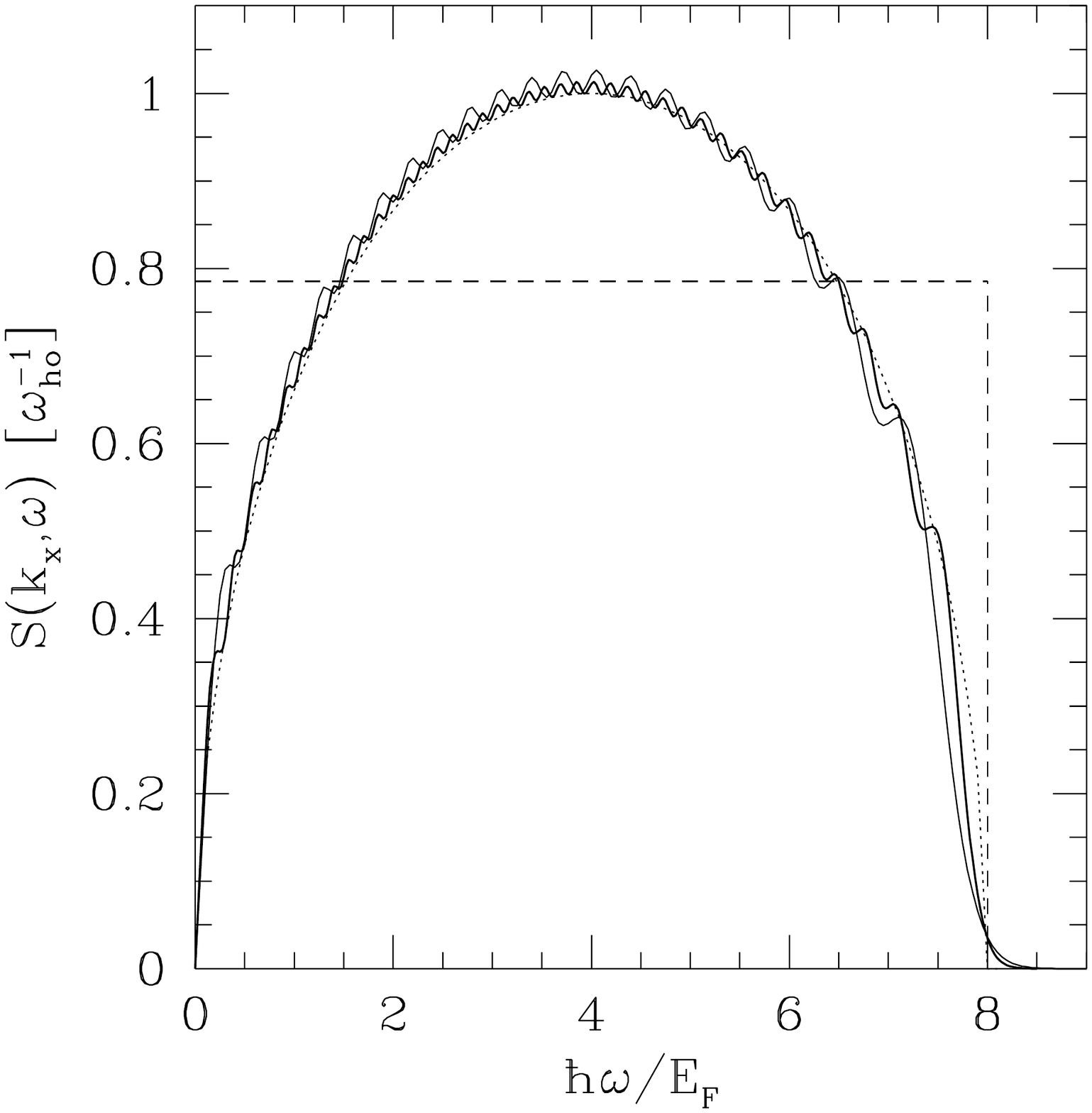}
    \caption{From [\onlinecite{Vignolo01}]. Dynamic structure factor of a TG gas under harmonic confinement. Left panel $k=0.1 k_F$, right panel $k=2 k_F$. The local density approximation (dots) is compared with the exact solution (lines) at various particle numbers. The dashed lines indicate the corresponding dynamic structure factor of a homogeneous gas.
    Reprinted figure with permission from  [\onlinecite{Vignolo01}],  \href{https://doi.org/10.1103/PhysRevA.64.023421}{https://doi.org/10.1103/PhysRevA.64.023421}. Copyright (2021) by the American Physical Society.}
    \label{fig:skw}
\end{figure}

As a first  illustration we present in Fig.~\ref{fig:skw} the results for the dynamical structure factor of a bosonic TG gas under harmonic confinement. Comparison with LDA shows that it provides a very good description of the spectrum if the energy transfer considered is much larger than the energy-level spacings $\omega_{0}$. The LDA approach has been used to predict the dynamical structure factors of bosons in a lattice plus harmonic trap confinement \cite{Golovach09}.
The presence of the harmonic trap provides important  qualitative changes in the shape of the dynamic structure factor as compared to the homogeneous case: in particular, due to the inhomogeneous density, the finite-$k$ small-$\omega$ regions are in this case accessible to excitations.

The exact solution for the dynamical structure factor has been extended at finite temperature \cite{lang2015dynamic}. As main effect of temperature, the backscattering region is washed out and higher energy excitations become possible. 

The dynamical structure factor can be also defined with respect to a non-equilibrium steady state: in this case its shape is considerably changed as compared to the ground-state one, reflecting the exotic nature of such state \cite{Denardis}.

\begin{figure}
    \centering
    \includegraphics[width=0.65\linewidth]{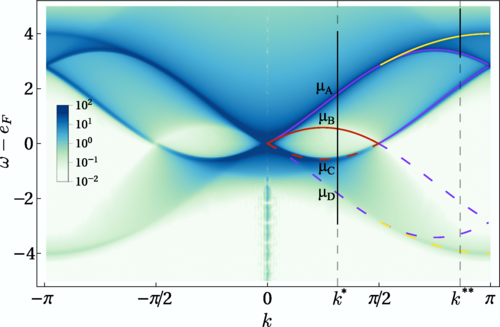}
    \caption{From [\onlinecite{settino2021exact}]. Spectral function of a TG gas on a lattice. The lines indicate the positions of the analogs of Lieb-I (red), Lieb-II (blue) branches and the third branch (yellow) typical of the lattice dispersion.
     Reprinted figure with permission from  [\onlinecite{settino2021exact}],  \href{https://doi.org/10.1103/PhysRevLett.126.065301}{https://doi.org/10.1103/PhysRevLett.126.065301}. Copyright (2021) by the American Physical Society.}
    \label{fig:akw}
\end{figure}

As a second example, we present the results for the  spectral function of a Tonks-Girardeau gas on a lattice \cite{settino2021exact}. Also in this case, the system is integrable only in the TG limit due to the presence of the external lattice potential and is shown in Fig.~\ref{fig:akw}. The spectral function contains three main excitation singularities: two of them related to the corresponding branches in the homogeneous system, namely the Lieb-I and Lieb-II branches, and a third one appearing only in lattices and associated to the the existence of an inflection point in the single-particle dispersion. It is interesting to notice that the Lieb-II branch, which has vanishing spectral weight in the dynamical structure factor, has here a diverging singularity. The measurement of the spectral function could then allow to observe for the first time this eluding branch.

\subsection{Momentum distribution}
\begin{figure}
\begin{center}
\includegraphics[width=0.65\linewidth]{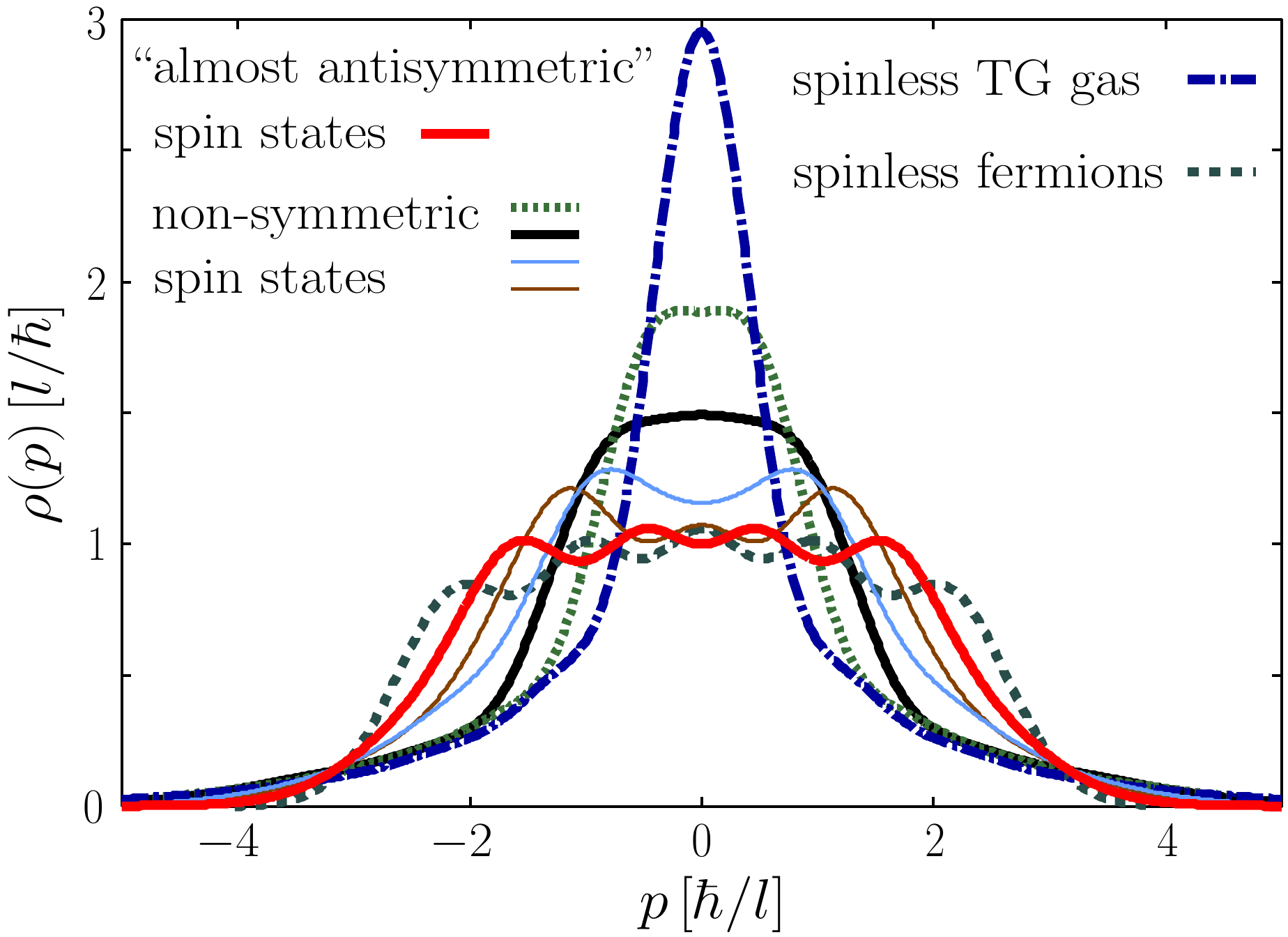}
\end{center}
\caption{\label{figBB-mom}From [\onlinecite{Deuretzbacher}].  Momentum distribution of 5 spin-1 bosons in different symmetry configurations in comparison with the momentum distribution of 5 spinless fermions.
 Reprinted figure with permission from  [\onlinecite{Deuretzbacher}],  \href{https://doi.org/10.1103/PhysRevLett.100.160405}{https://doi.org/10.1103/PhysRevLett.100.160405}. Copyright (2021) by the American Physical Society.
}
\end{figure}

\begin{figure}
\begin{center}
\includegraphics[width=0.65\linewidth]{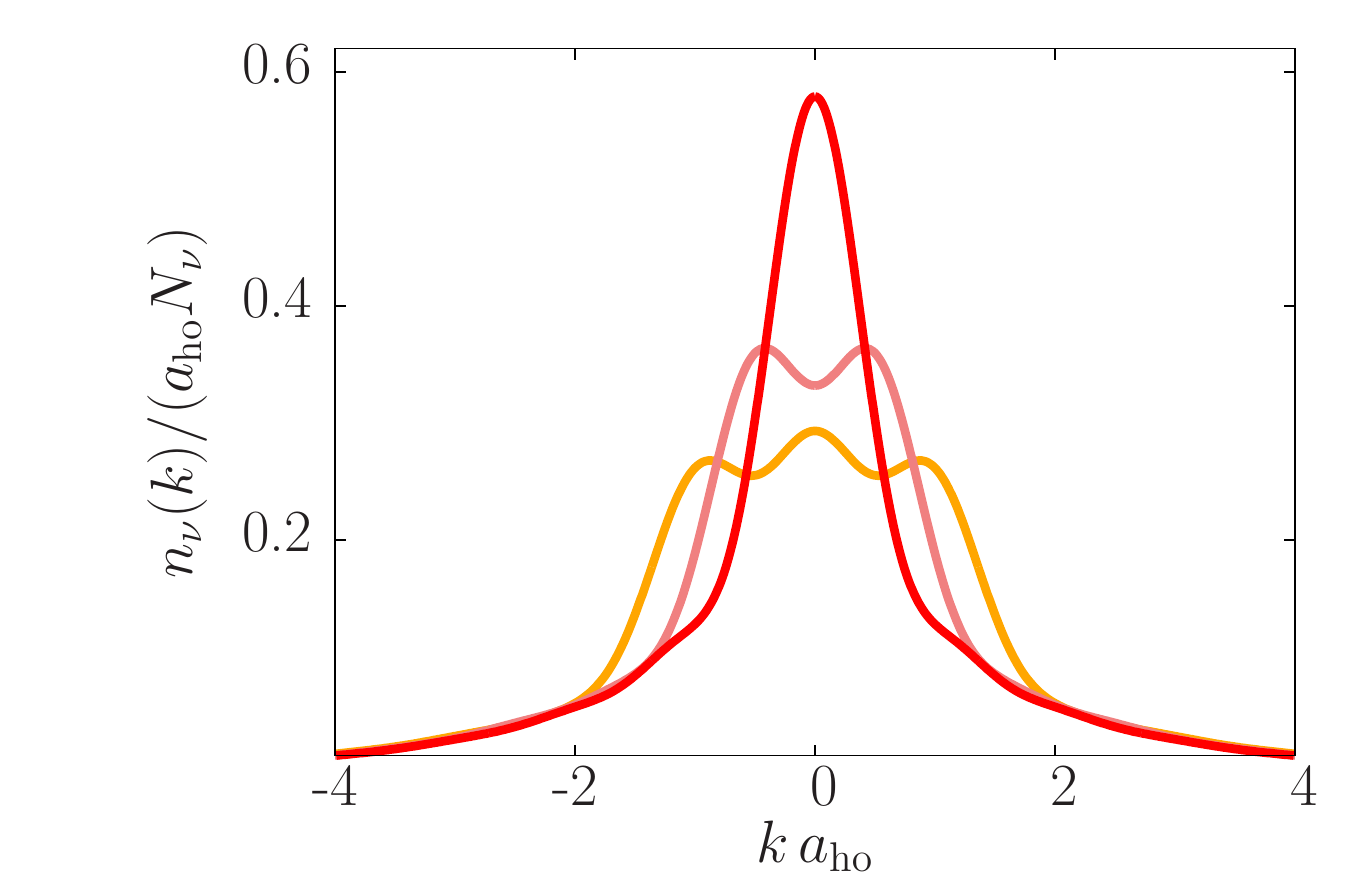}
\end{center}
\caption{\label{figFF-mom}From [\onlinecite{Decamp2016-2}].  Momentum distribution for a balanced 6-component (red curve), 3-component (pink curve) and 2-component (orange curve) fermionic mixture with a total number of 6 fermions. Reprinted figure with permission from  [\onlinecite{Decamp2016-2}],  \href{https://doi.org/10.1103/PhysRevA.94.053614}{https://doi.org/10.1103/PhysRevA.94.053614}. Copyright (2021) by the American Physical Society.}
\end{figure}

\begin{figure}
\begin{center}
\includegraphics[width=0.63\linewidth]{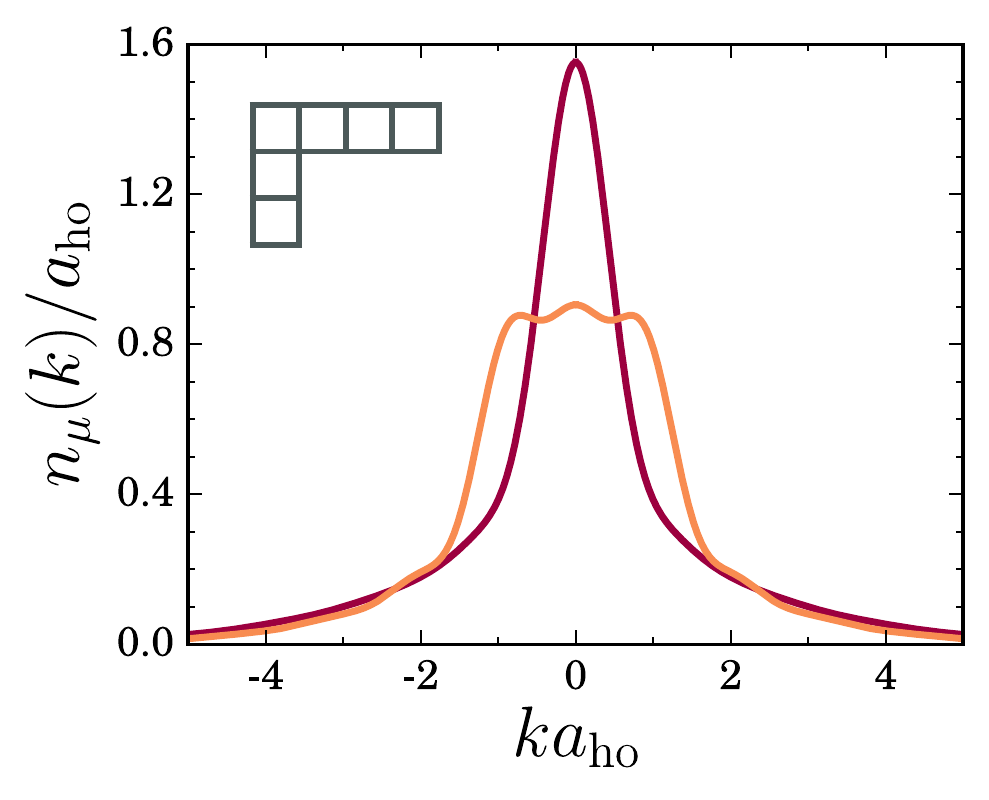}
\end{center}
\caption{\label{figBF-mom}From [\onlinecite{Decamp2017}].  Momentum distribution for a mixture of 3 bosons and 3 fermions. The maroon line shows the bosonic momentum distribution and the orange line, the fermionic one. Reprinted with permission from Decamp {\it et al.}, New J. Phys. {\bf 19},  125001  (2017). Copyright 2017, (https://doi.org/10.1088/1367-2630/aa94ef). Author(s) licensed under a Creative Commons Attribution 4.0 License.}
\end{figure}

The momentum distribution of a harmonically trapped TG gas was obtained in Ref.~[\onlinecite{GirWriTri01}]. 
An analytical closed formula for the momentum distribution of two TG bosons is also known \cite{Ancarani2021}. Since the momentum distribution is an off-diagonal observable, ie related to the one-body density matrix,  its shape is different from the one of  a spinless fermionic gas, which, in harmonic trap, coincides with the density profile. The momentum distribution  of the TG gas displays a a unique central peak scaling with $\sqrt{N}$ and algebraic tails at large momenta. Also, we notice that there are no oscillations.
This is shown in Fig.~\ref{figBB-mom}. The authors of this work\cite{Deuretzbacher} have calculated the momentum distribution for 5 spin-1 bosons, for different wavefunction symmetries and have compared them to the one for 5 spinless fermions. 
The fully symmetric state corresponds to the spinless TG gas. For less symmetric states, the peak splits,
and the momentum distribution develops oscillations, the more the state is anti-symmetric.
The momentum distribution for particles with exchange symmetry depends on the allowed symmetry and not really on the nature of the particles themselves\cite{Dehkharghani2015}. As an illustration of this idea, we show in Figs.~\ref{figFF-mom} and \ref{figBF-mom} the momentum distribution for 6 particles. Fig.~\ref{figFF-mom} corresponds to various choices of balanced fermionic mixtures whose density profiles are shown in Fig.~\ref{figFF-dens}.  Fig.~\ref{figBF-mom} refers to the ground-state for 3 spinless fermions and 3 identical bosons.
The case of 6-component 6 fermions corresponds to a fully symmetric wavefunction and one finds a momentum distribution identical to that of a TG gas. For less symmetric cases, as for the spin-1 bosons, the momentum distribution develops a number of momentum density oscillations equal to the length of the longest column
of the corresponding Young tableaux: 2 for the case of a 3-component 6-fermion mixture, 3 for the two-component case and for the 3-bosons-3-fermions mixture.

The effect of an impurity in a trapped Bose system with different mass ratio has been studied in [\onlinecite{Dehkharghani2015b}].

At finite temperature, the momentum distribution of multicomponent mixtures displays a crossover behaviour as a function of temperature $T$, going from a 'spin-ordered' state at low temperature to a 'spin-disordered' one at high temperature \cite{Cheianov2005}. The crossover occurs when $k_B T$    exceeds the energy difference among energy levels within the ground state manifold, hence the typical crossover temperature scales as $1/g$.

\subsection{Tan's contact}
\begin{figure}
\begin{center}
\includegraphics[width=0.65\linewidth]{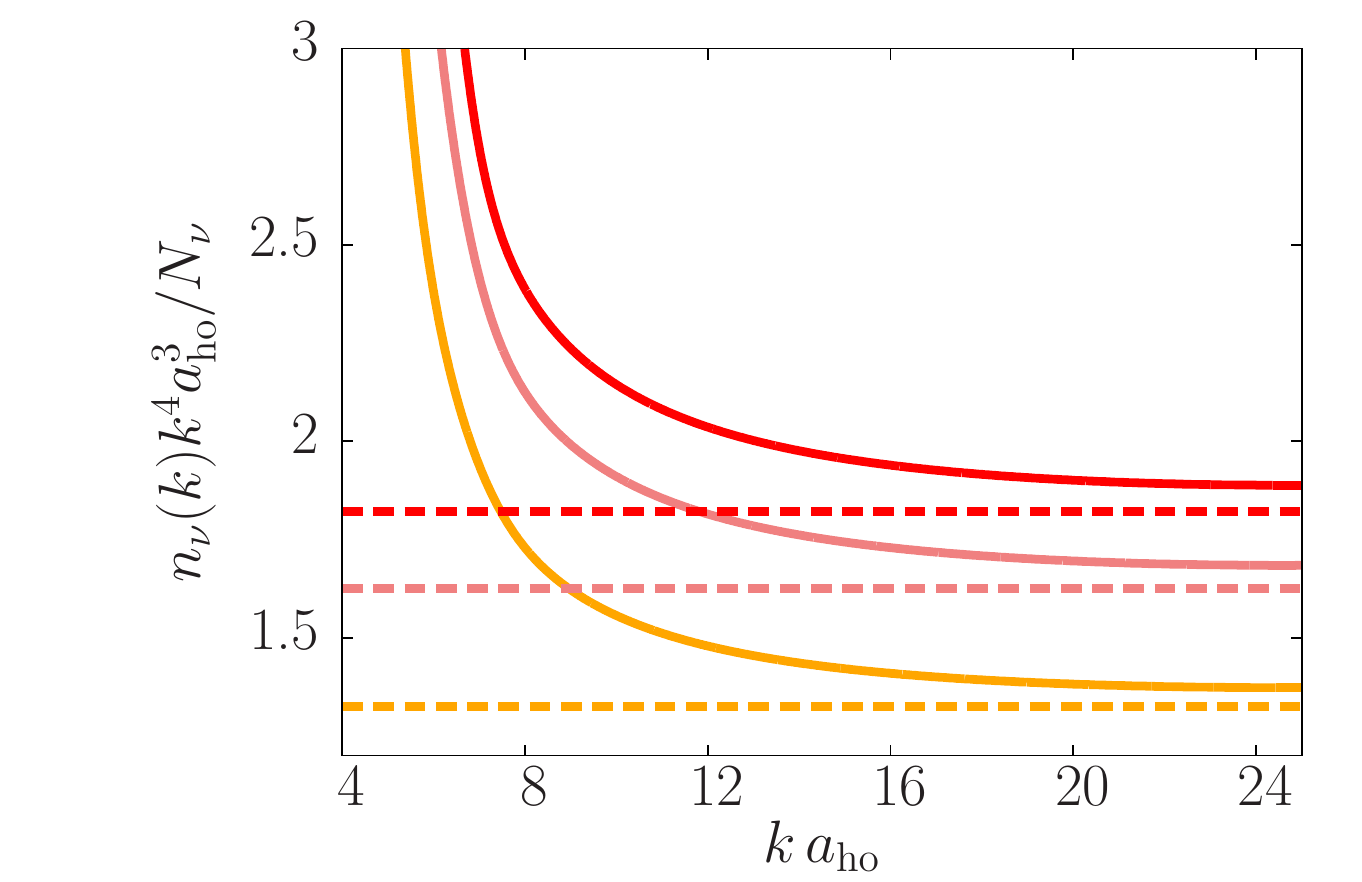}
\end{center}
\caption{\label{figFF-cont}From [\onlinecite{Decamp2016-2}].  Tan's contact ($n(k)k^4$) for a balanced 6-component (red curve), 3-component (pink curve) and 2-component (orange curve) fermionic mixture with a total number of 6 fermions. The data are the same of those of Fig.~\ref{figFF-mom}.The asymptotic values (horizontal lines) have been evaluated from Eq.~(\ref{formula-c-jean}).
Reprinted figure with permission from  [\onlinecite{Decamp2016-2}],  \href{https://doi.org/10.1103/PhysRevA.94.053614}{https://doi.org/10.1103/PhysRevA.94.053614}. Copyright (2021) by the American Physical Society.}
\end{figure}

As the momentum distribution depends on the wavefunction symmetry, the Tan's contact depends on it as well. In some way, the contact counts the many-body wavefunction cusps, thus more the wavefunction is symmetric, more the contact is sizeable. In the opposite situation, for a fully anti-symmetric state, the Tan's contact will be zero.
In Fig.~\ref{figFF-cont} we show the tails of the momentum distributions, drawn in Fig.~\ref{figFF-mom}, multiplied by $k^4$: the asymptotic value at large $k$ gives the Tan's contact. The largest contact corresponds to the fully-symmetric wavefunction (6 fermions, 6 components), the second to the two-rows diagram {\tiny\yng(3,3)} (6 fermions, 3 components) and the last to the diagram {\tiny\yng(2,2,2)} (6 fermions, 2 components).
The Tan's contact is thus the fingerprint of the many-body wavefunction symmetry. 
Let us remark that the Tan's contact can be exactly calculated for two harmonically trapped bosons at any interaction strength\cite{Rizzi2018},
\begin{equation}
\mathcal{C}_2(g)=\dfrac{m^2g^2}{\pi\hbar^4} |\psi_\nu(0)|^2,
  \label{eq:c2}
\end{equation}
$\psi_\nu(x_{rel})$ being given in Eq.~(\ref{rizzi}).
Eq.~(\ref{eq:c2}) provides the well-known limit for the TG gas\cite{Santana2019}: $\mathcal{C}_2(\infty)=(2/\pi)^{3/2}a_{ho}^{-3}$.

\subsection{Dynamical properties and quenches}
\label{Sec:quenches}

One strength point of the TG solution is the possibility to describe the arbitrary quantum dynamics, including situations strongly out-of equilibrium.

As first example, we describe the dynamics of a TG gas following  a sudden turn-off of the harmonic confinement $V_{ext}(x,t)=m\omega^2(t) x^2/2$ with $\omega(t)=\omega_0$ for $t<0$ and $\omega(t)=0$ for $t\ge 0$. Notice that there is no expansion in the transverse direction, the motion corresponds to the expansion inside a one-dimensional waveguide. For this reason, at difference from the usual three-dimensional expansion, interactions during expansion cannot be neglected and indeed strongly influence the dynamics. 

\begin{figure}[tb]
\centerline{\includegraphics[width=7.5cm,angle=270]{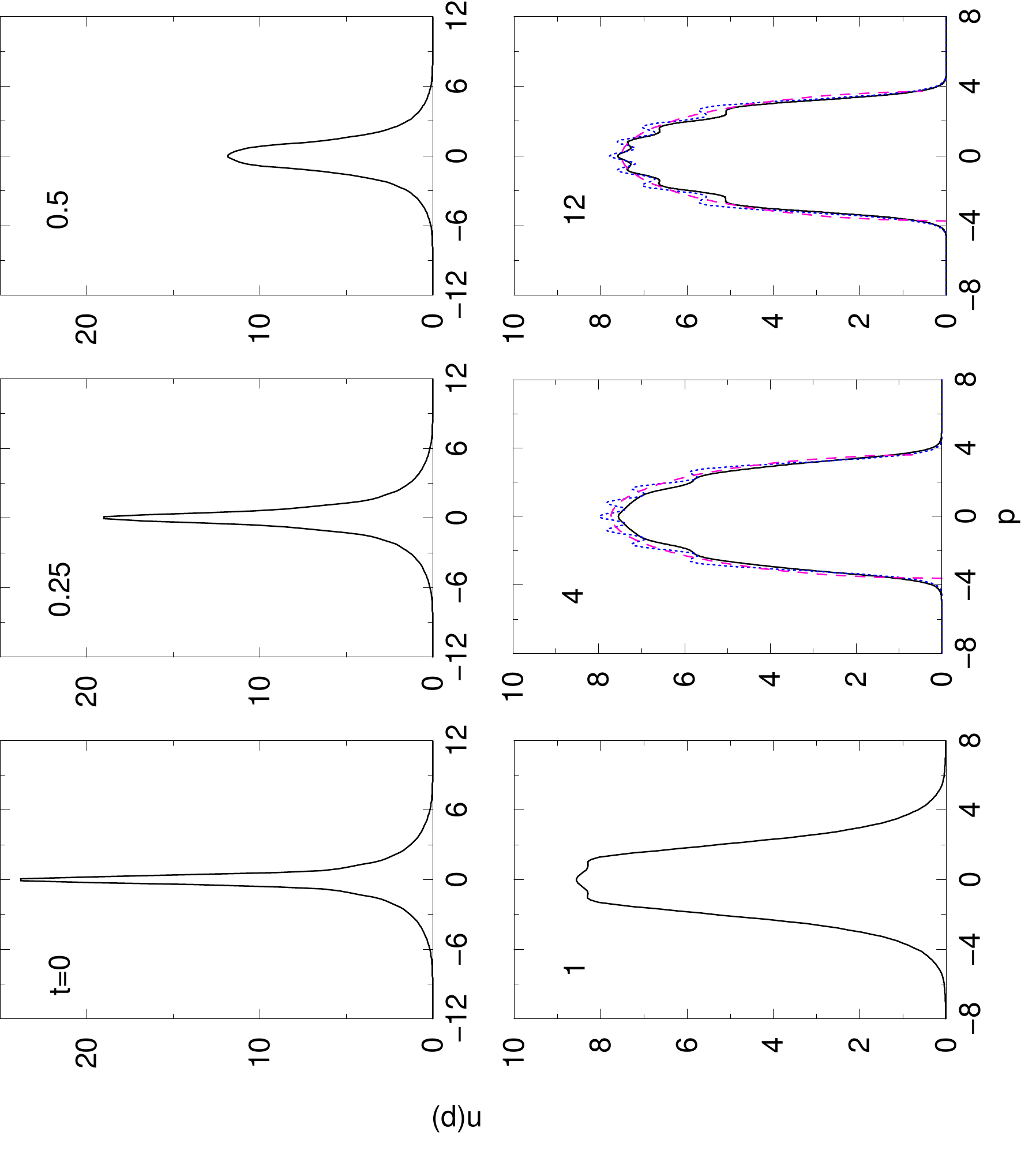}} 
\vspace{0.2cm}
\caption[]{From  [\onlinecite{Minguzzi05}]. Momentum distribution (in units of $\hbar/a_{ho}$) as a function of time (in units of $\omega_{0}^{-1}$) of an expanding TG gas following a sudden turn-off of the harmonic potential. 
At long times, the momentum distribution tends to the one of an equilibrium Fermi gas under harmonic confinement. 
Reprinted figure with permission from  [\onlinecite{Minguzzi05}],  \href{https://doi.org/10.1103/PhysRevLett.94.240404}{https://doi.org/10.1103/PhysRevLett.94.240404}. Copyright (2021) by the American Physical Society.
}
\label{fig:tgexpans}
\end{figure}

To describe the dynamics we use the time-dependent Bose-Fermi mapping (see Sec.\ref{sec:exact-sol-bosons}). The specific expansion dynamics can be solved exactly at arbitrary times \cite{Minguzzi05} by introducing a scaling parameter $b(t)=\sqrt{1+\omega_{0}^2t^2}$ associated to the size of the density profile during the expansion and a dynamical phase. The TG wavefunction is then expressed in terms of the one at initial times according to 
\begin{eqnarray}
  \label{eq:scaling_mb}
  \Phi_B(x_1,..,x_N;t) = b^{-N/2} \Phi_B(x_1/b,..,x_N/b;0) \nonumber \\ \times 
  \exp\left(\frac{i\dot{b}}{b\omega_0}\sum_j
  \frac{x_j^2}{2a^2_{ho}}\right)\exp\left(-i \sum_j \epsilon_j \tau\right).
  \end{eqnarray}
This solution allows to calculate several properties, such as the time-dependent density profile and the momentum distribution. A remarkable prediction stemming from the above solution is that the momentum distribution at long times tends to the one of a non-interacting Fermi gas ("dynamical fermionization") as also  observed in numerical simulations on a lattice \cite{Rigol2005} and  experimentally \cite{Wilson2020}.  The state of the system at such long times is well described by generalized Gibbs Ensemble and the equilibration mechanism has been elucidated in terms of interference effects \cite{collura2013equilibration,collura2013quench}.

The same type of solution describes also  a partial opening of the trap, described by $\omega(t\le
0)=\omega_{0}$ and $\omega(t)=\omega_1$ for $t>0$. This excites a  large-amplitude breathing mode. Correspondingly, the momentum distribution oscillates in time between the one of a TG gas and a fermionic one \cite{Minguzzi05}. A remarkable feature of the oscillation is that it is undamped. This is related to integrability and constrained dynamics in one dimension.
At finite temperature the exact TG solution predicts a many-body bounce effect \cite{atas2017collective}, ie the narrowing of the momentum distribution  at twice the rate of oscillations of the density profile. Frequency doubling in momentum space was also experimentally observed at weak interactions \cite{fang2014quench-induced}.
At strong finite interactions, no exact theory is available but Generalized Hydrodynamics \cite{castro-alvaredo2016emergent,bertini2016transport} well accounts for the quench dynamics observed in the experiment \cite{malvania2021generalized}.

The role of confining potential in the quench dynamics has also been explored.  The release of a TG gas from a hard wall trap was studied, and a notable difference was found on the scaling of the thermalization time with particle number \cite{delcampo2006dynamics}. 
The dynamics of a TG gas following a sudden change of trap strength for a quartic potential was addressed in Ref.[\onlinecite{fogarty2020manybody}], showing that interparticle collisions allow the TG gas to decohere more quickly than a non-interacting Fermi gas, due to different properties of the off-diagonal part of the one-body density matrix of the two gases. The same work also pointed out the different dynamical behaviour of TG bosons and ideal fermions in  shortcut-to-adiabaticity protocols.

The exact solution for the arbitrary quantum dynamics can be used to access to a wealth of dynamical problems and regimes. For example, it is possible to follow the dynamics of a TG gas in presence of a barrier potential. Ref.[\onlinecite{goold2010eccentrically}] reports of the equivalent of the optical Talbot effect in the dynamics following the sudden turn off of an eccentric barrier potential. Barrier renormalization effects due to quantum fluctuations can be probed by the time evolution following sudden displacement of a harmonic trap split by a barrier \cite{cominotti2015dipole}. 
The dynamics of population imbalance across a barrier allows to follow the Josephson oscillations among tunnel-coupled one-dimensional tubes in a head-to-tail configuration. The exact TG solution has provided a stringent test of the Luttinger-liquid theory predictions and highlighted some low-energy excitation modes responsible for the damping of the Josephson oscillations \cite{polo2018damping}. 

Putting a TG gas on a ring, one can follow the current flows. Also in this case, important information can be obtained from the exact TG solution. For example, it has been shown that in presence of a weak barrier potential, an initially phase imprinted current undergoes coherent oscillations, ie it displays quantum coherent phase slips. The TG solution allows to access to the nature of the state during the dynamics and show that multi particle-hole oscillations play a major role \cite{polo2019oscillations}. 

Large-amplitude quench dynamics can also be engineered to give rise to dispersive shock waves in TG gases. Two protocols have been proposed: i) a sudden change of a localised external potential \cite{damski2004shock, simmons2020what}  and ii) a quantum fluid  hitting against the hard walls of its container \cite{dubessy2021universal}. 

A phenomenological model for the dynamics of an output-coupled TG gas traversing its parent cloud as the one experimentally realized in Ref.~[\onlinecite{Palzer09}] was proposed in Ref.~[\onlinecite{rutherford2011transport}].

Other types of quenches allow to study phase transitions, as eg it is the case for the pinning and commensurate-incommensurate transition in the presence of an optical lattice. A sudden turn off of the optical lattice gives rise to a dynamical depinning of the TG gas \cite{cartarius2015dynamical} and the sudden set into motion of the lattice allows to probe the various phases \cite{mikkelsen2018static}.  

The quench dynamics of fermionic gases with strong repulsions has also attracted some attention. The dynamics of SU$(\kappa)$ fermions following a sudden change of the trapping potential was addressed in Ref.~[\onlinecite{barfknecht2019dynamics}], finding a decoupling of density and spin dynamics and suppression of the latter.  Ref.~[\onlinecite{pecci2021universal}] proposes a setup to observe the spin mixing dynamics following an initially fully imbalanced state, and finds universal oscillations and superdiffusion magnetization dynamics. This last result, already predicted\cite{ljubotina2017spin} and observed\cite{wei2021quantum} for a 1D homogeneous spin system, shows that superdiffusion persists also in the presence of an external potential that breaks the integrability of the system.


\subsection{Finite temperature results}
\begin{figure}
\begin{center}
\includegraphics[height=0.49\linewidth,angle=270]{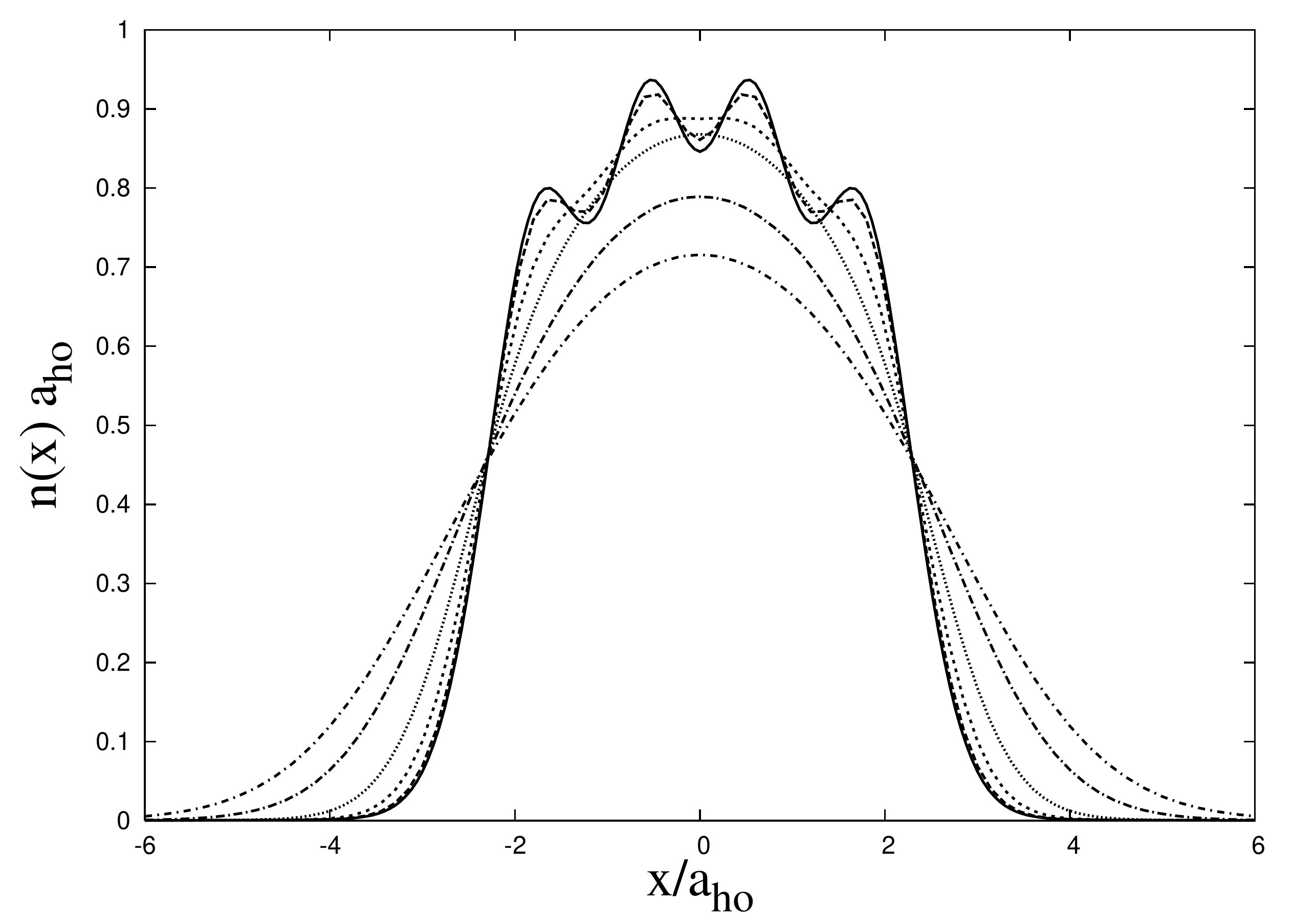}
\includegraphics[height=0.49\linewidth,angle=270]{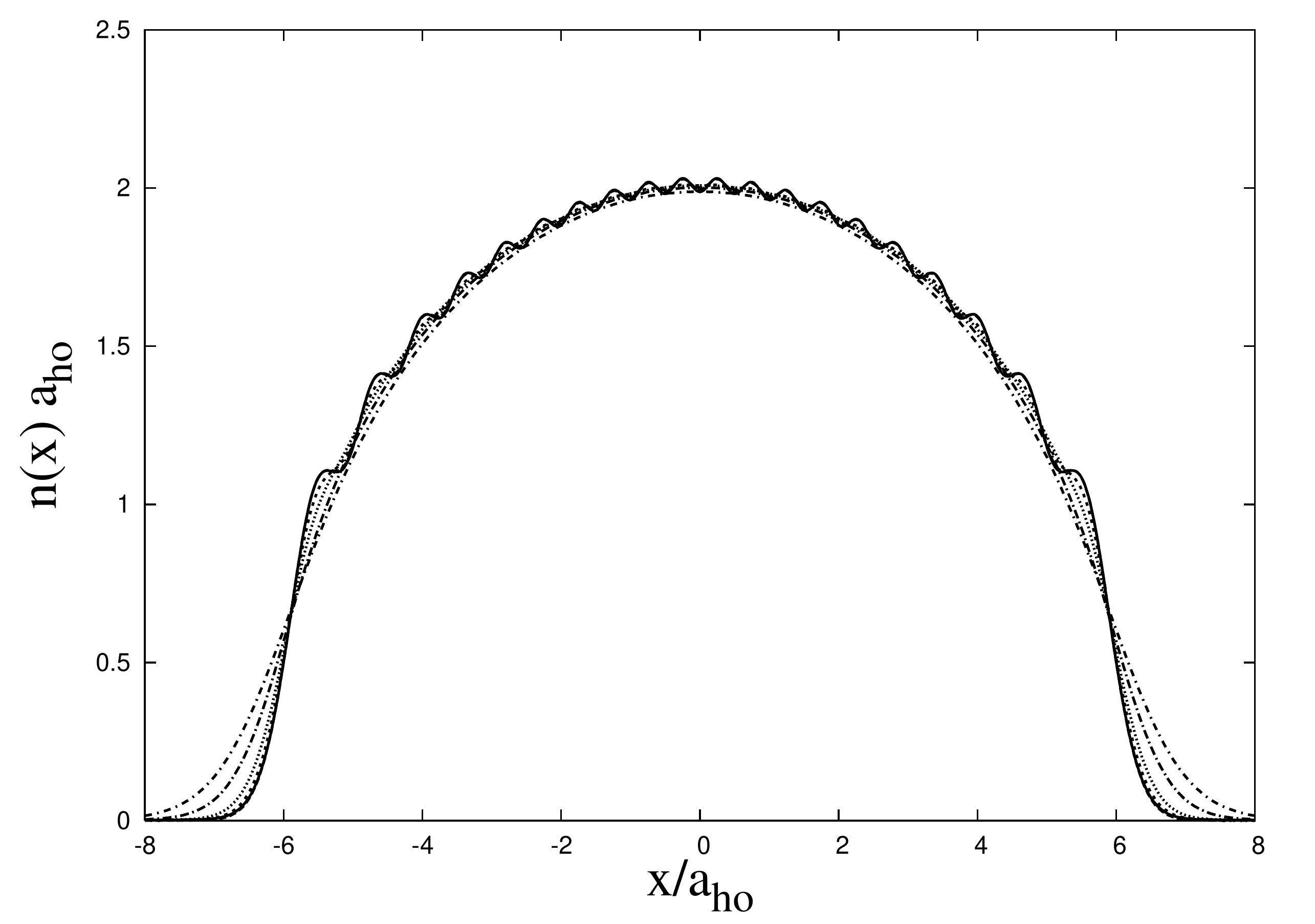}
\end{center}
\caption{\label{fig-zerha-T}From [\onlinecite{Akdeniz2002}].  Particle density profile for $N = 4$
(left panel) and $N=20$ (right panel) harmonically confined TG bosons (or non-interacting fermions) at various values of the temperature. 
$ T = 0$ (solid curve) and $T = 0.2 \hbar\omega_0/k_B$ (dashed curve); the other
curves refer to $k_BT /\hbar\omega_0$ = 0.5, 1.0, 2.0 and 3.0, in order of decreasing peak height.
Reprinted figure with permission from  [\onlinecite{Akdeniz2002}], \href{https://doi.org/10.1103/PhysRevA.66.055601}{https://doi.org/10.1103/PhysRevA.66.055601}. Copyright (2021) by the American Physical Society.}
\end{figure}

\begin{figure}
\begin{center}
\includegraphics[width=0.65\linewidth]{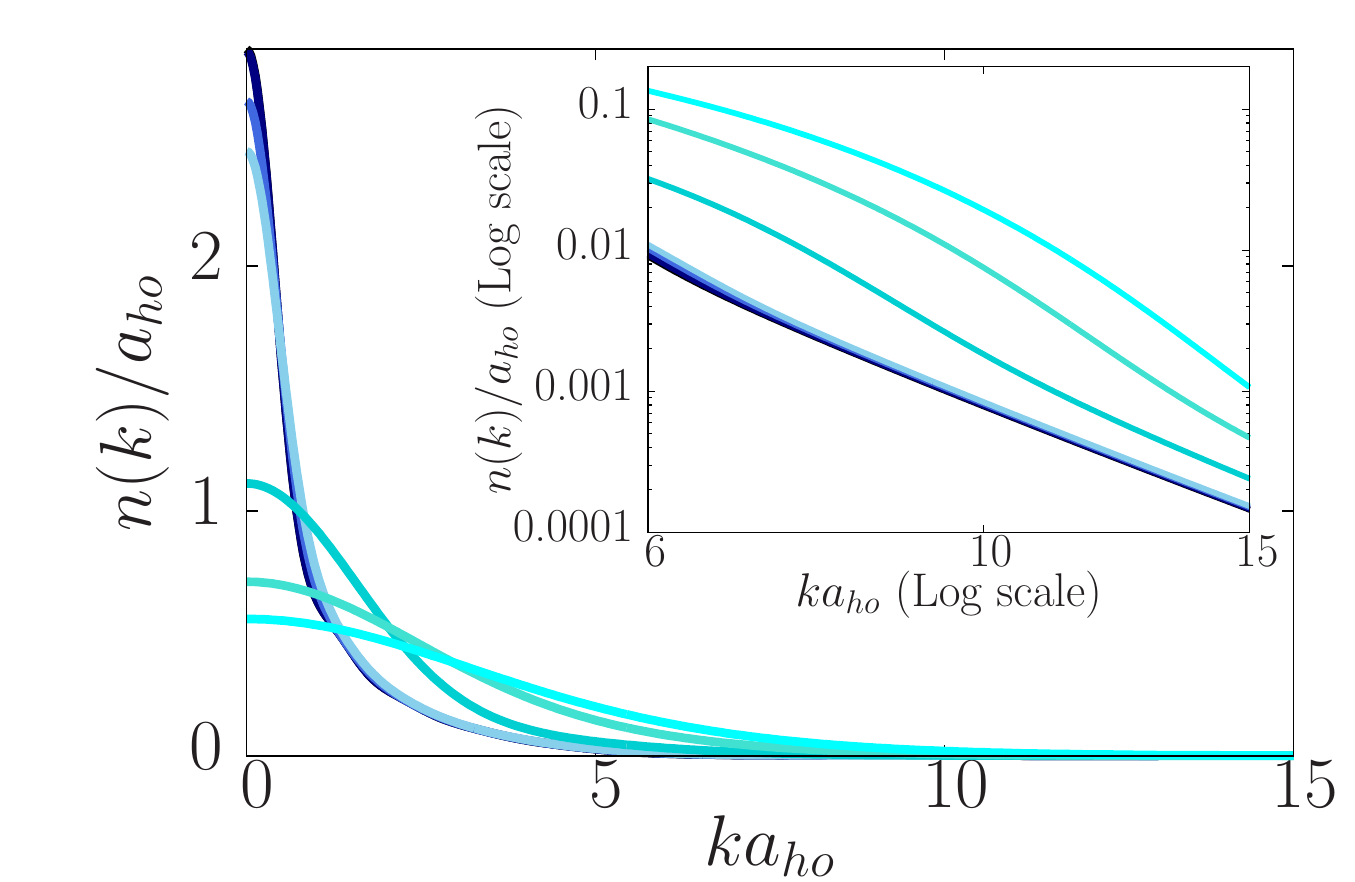}
\includegraphics[width=0.65\linewidth]{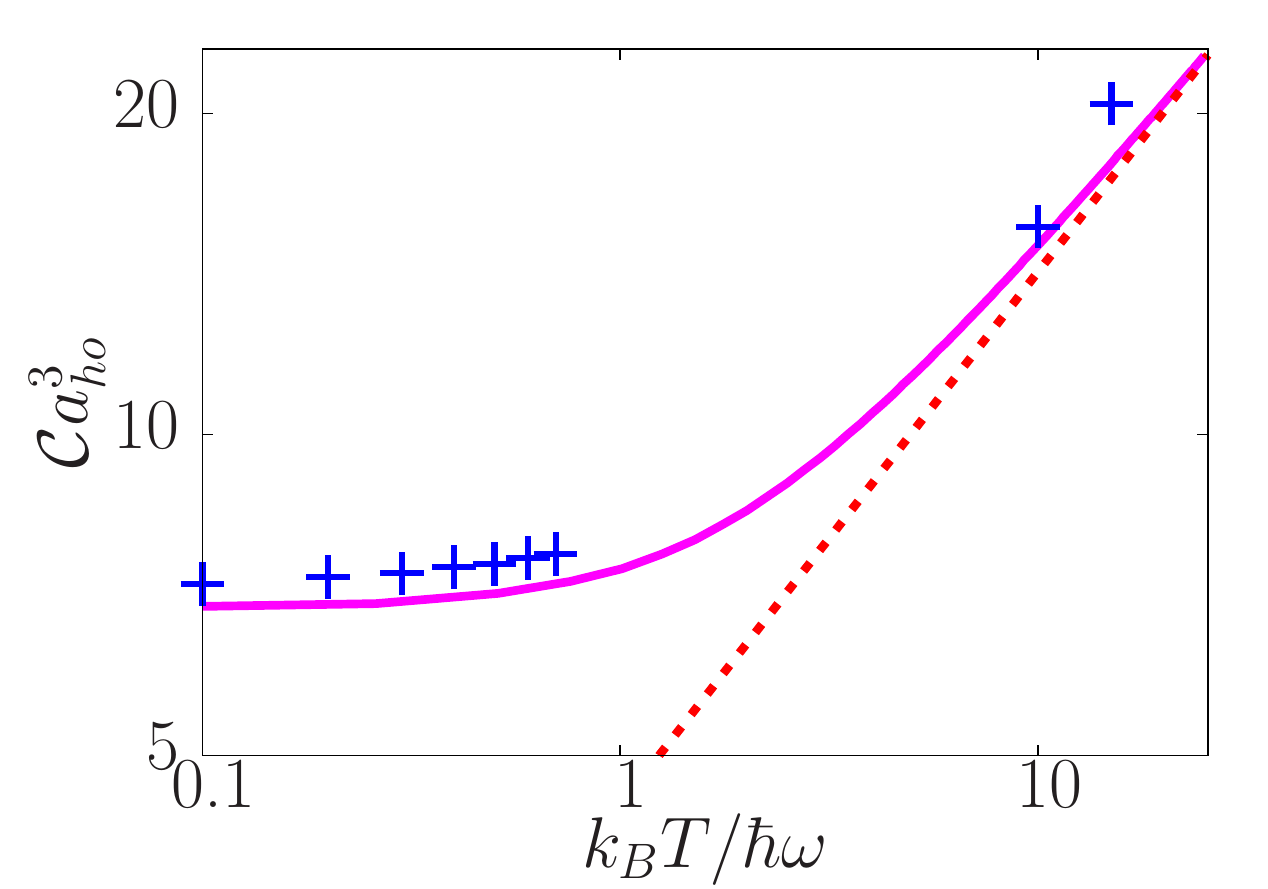}
\end{center}
\caption{\label{fig-PRL-AP}From [\onlinecite{vignolo2013}]. Top panel: momentum distribution of a TG gas (in units of $a_{ho}$) as a function of wavevector (in
units of $1/a_{ho}$) with $N = 5$ particles under harmonic confinement at increasing temperature, from top to bottom in the
main peak $k_BT /\hbar\omega_0$=0.1, 0.5, 0.7, 5, 10, 15. 
Bottom panel: Tan's contact (in units of $a^3_{ho}$) as a
function of temperature $k_BT$ (in units of $\hbar\omega_0$) for a TG gas under harmonic confinement. The expression from (\ref{bess}) and (solid, magenta) is compared with the high-temperature
limit (\ref{virial-TG}) (dashed, red) and the data from the numerical calculation of the momentum distribution (crosses, blue).
Reprinted figure with permission from  [\onlinecite{vignolo2013}], \href{https://doi.org/10.1103/PhysRevLett.110.020403}{https://doi.org/10.1103/PhysRevLett.110.020403}. Copyright (2021) by the American Physical Society.
}
\end{figure}

Signatures of quantum correlation, as the shell effects  in the density profiles, and in the bulk of the momentum distribution are washed out already at temperatures of the order of the harmonic oscillator energy spacing $T\sim \hbar\omega_0/k_B$.
This is shown in Fig.~\ref{fig-zerha-T} for the density profiles of $N=4$ and 20 TG bosons
\cite{Akdeniz2002} and in the first panel of Fig.~\ref{fig-PRL-AP} for the momentum distribution of 5 TG bosons at increasing temperatures.

However, the contact in 1D is surprisingly robust against temperature. It is even better: in the TG regime it increases with the temperature\cite{vignolo2013}, as we have shown in Eq.~(\ref{virial-TG}). This counter-intuitive
result is shown  in the inset of top panel of Fig.~\ref{fig-PRL-AP} and in the bottom panel of the same figure.
The temperature does not wash out the cusps in the hard-core limit and moreover allows the particles to get closer, increasing the slope of the wavefunction in the neighbourhood of the cusps, thus increasing the contact.

At finite interaction, in the harmonically trapped system, the contact first increases with
the temperature till the value $T_{max}\simeq mg^2/(8\hbar^2k_B)$ and then decreases\cite{Yao2018}. This can be deduced from the virial expression (\ref{virial-mia}). The maximum marks the crossover between the fermionized regime and the ideal boson gas. In the TG regime the position of this maximum tends to infinity (\ref{virial-TG})
since fermionization persists at any temperature.

\begin{figure}
\begin{center}
\includegraphics[width=0.65\linewidth]{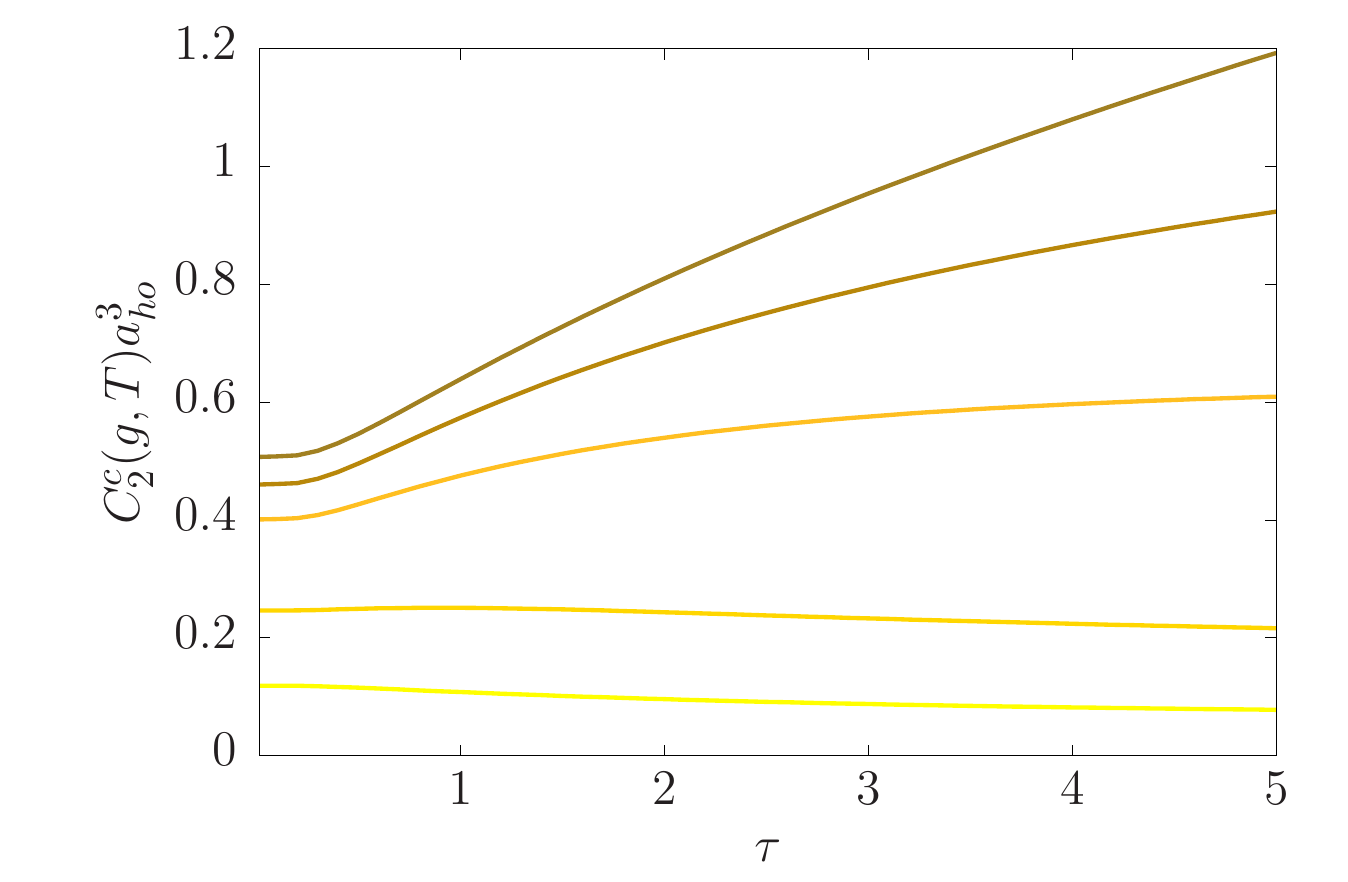}
\end{center}
\caption{\label{fig-can-T}From [\onlinecite{Santana2019}]. Canonical Tan's contact $C^c_2(g,T)$ as a function of $\tau=T/T_F$ for different values of the interaction strength $z=a_{ho}/(|a_{1D}|\sqrt{N})$. From bottom to top: $z=0.5$, 1, 2.5, 5, and 1000. The curve for $z=1000$ is indiscernable from the contact evaluated in the TG limit (\ref{two-bos-inf}).
Reprinted figure with permission from  [\onlinecite{Santana2019}], \href{https://doi.org/10.1103/PhysRevA.100.063608}{https://doi.org/10.1103/PhysRevA.100.063608}. Copyright (2021) by the American Physical Society.
}
\end{figure}

At finite interactions and finite temperature, the contact for a harmonically trapped system can be exactly calculated only for two particles, since the whole spectrum is known\cite{Busch98}.
An analytical expression can be derived for the case of two TG bosons (in the canonical ensemble). It reads
\begin{equation}
  \mathcal{C}_2^c(\infty,T)=\frac{\sqrt{32}}{\pi^{3/2} a_{ho}^{3}} Z_r^{-1} \sum_{j}e^{-\beta \hbar\omega_0 (2j-1)} \dfrac{(2j-1)!!}{2^j(j-1)!},
  \label{two-bos-inf}
\end{equation} 
with $Z_r=\sum_{j}e^{-\beta\hbar\omega_0(2j-1)}$. One can readily check that  the zero-temperature limit of Eq.~(\ref{two-bos-inf}) yields $\mathcal{C}_2(\infty,0)=(2/\pi)^{3/2}a_{ho}^3$.
The results for the contact of two bosons as a function of the temperature  for various values of interaction strength are shown in Fig.~\ref{fig-can-T}.
We will see in the next section that the two-body calculation encloses an essential part of the
contact for $N$ particles.
\subsection{Scaling properties}
In the regimes discussed in Sec. \ref{two-is-enough}, where Eq.~(\ref{miamia}) holds for any $N$, we can write
\begin{equation}
{\mathcal{C}_N(\xi_\gamma,\tau)}={\mathcal{C}_2(\xi_\gamma,\tau)}
\dfrac{\mathcal{C}_N(\infty,\tau)}{\mathcal{C}_2(\infty,\tau)}.
\label{miamia2}
\end{equation}
Eq.~(\ref{miamia2}) can be interpreted as follows: (i) at each temperature the way in which particles see each other at a given interaction strength is given
by the two-body calculation; (ii) the correlation contribution due to the fact that the particles are $N$ and not only two, is embedded in the contact at in the $g \rightarrow \infty$ limit calculated at the same temperature.
Let us underline that both the two-body contact at finite temperature and interaction strength, and the contact for $N$ TG particles at finite temperature can be calculated exactly. Moreover, for the canonical ensemble, that is the relevant case for experiments, it exists an analytical Ansatz for the finite-temperature TG gas contact\cite{Santana2019}
\begin{eqnarray}
  \mathcal{C}_N^c(\infty,\tau)&=&h_2(\infty,\tau)s(N) \label{bellissima}\\
  &=&h_2(\infty,\tau)\left(N^{5/2}-N^{3/4(1+\exp(-2/\tau))}\right),\nonumber
\end{eqnarray}
where
\begin{equation}
  h_2(\infty,\tau)=\mathcal{C}_2(\infty,T(\tau))/s(2).
\label{rottura}
\end{equation}
The function $s(N)$ interpolates between the ($N^{5/2}-N^{3/4}$) scaling at zero temperature\cite{Rizzi2018} and the ($N^{5/2}-N^{3/2}$) scaling for canonical ensembles at large temperature\cite{Santana2019}.
We expect Eq.~(\ref{miamia2}) to hold also for boson-boson and boson-fermion mixtures at zero temperature and at very large temperatures, both in the canonical and grand-canonical ensembles.

\begin{figure}
  \begin{center}
    \includegraphics[width=0.8\linewidth]{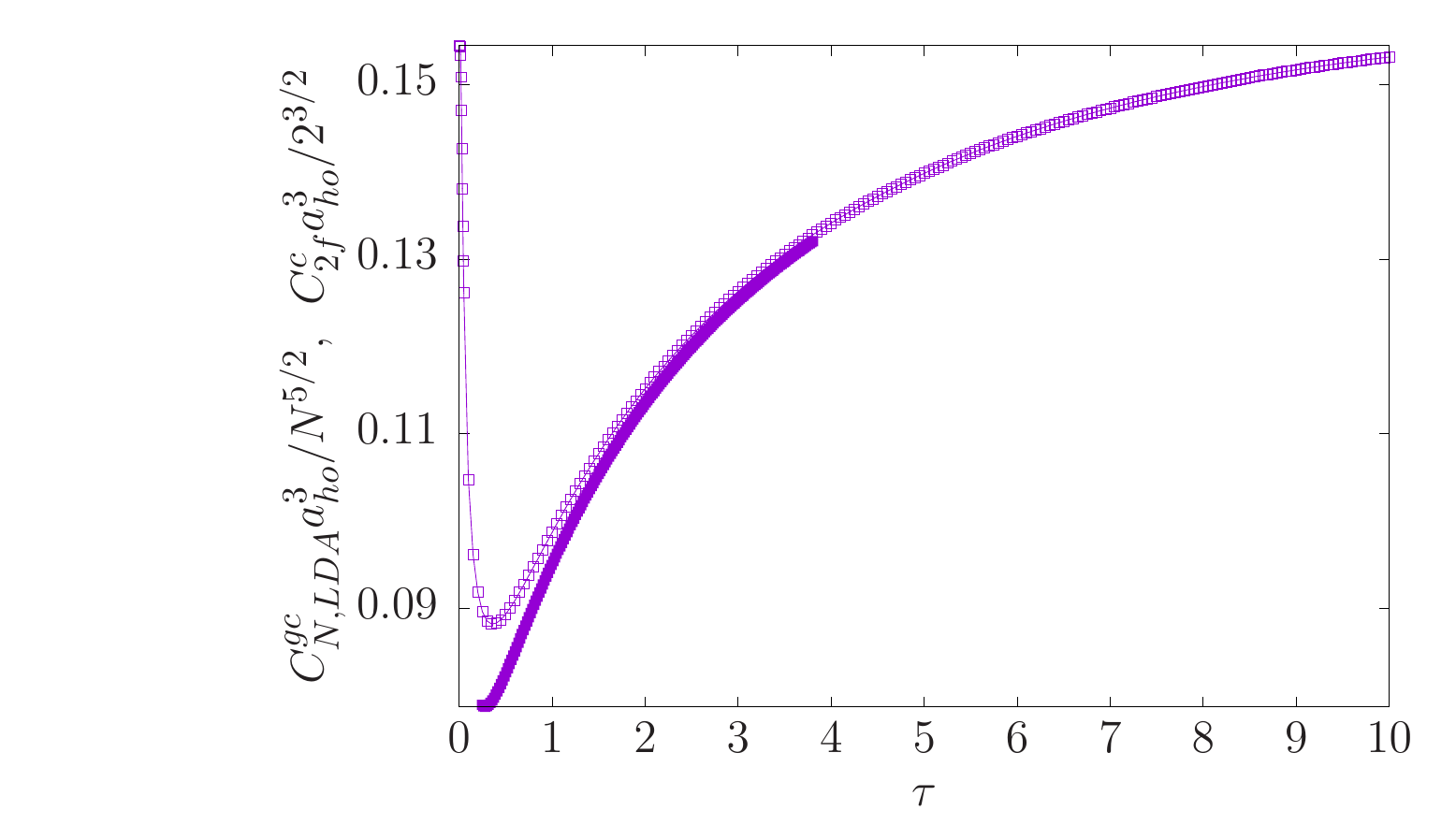}
     \end{center}
     \caption{From  [\onlinecite{Capuzzi2020}].
     LDA grand canonical contact $\mathcal{C}_{N,LDA}^{gc}$ rescaled
       by $N^{5/2}$ (full symbols) and the canonical one $\mathcal{C}_{2f}^c$ for two fermions
       (empty symbols) rescaled by $N^{3/2}(N-1)=2^{3/2}$ as functions
       of $\tau=T/T_F$, for the case
       $\xi_{\gamma}=3.53$.
   Reprinted figure with permission from  [\onlinecite{Capuzzi2020}], \href{https://doi.org/10.1103/PhysRevA.101.013633}{https://doi.org/10.1103/PhysRevA.101.013633}. Copyright (2021) by the American Physical Society.    
       \label{fig-su2-gc-c}}
\end{figure}

More challenging is the analysis of multi-component mixtures at finite, low temperatures\cite{Cheianov2005,Patu2016,Capuzzi2020}. In this regime there is a sort of symmetry mixing with differing weights depending on the temperature and the symmetry itself, that causes a rapidly drop of the contact with the temperature. The characteristic temperature $T_0$ of such a symmetry blending, for the trapped system and in the strong interacting limit, scales with the ground-state contact in the $g \rightarrow \infty$ limit divided by the interaction strength\cite{Capuzzi2020}. Thus, in such limit, the drop of the contact is a discontinuous jump
at $T=0$ ($T_0\rightarrow 0$). This means that, in this range of temperature, $\mathcal{C}_N(\infty,T)$ cannot catch
the $N$ dependence of $\mathcal{C}_N(g,T)$ at finite $g$, that is a continuous function of the temperature.
However,  for the case of a SU(2) fermionic mixture, by performing a LDA calculation on the top of the thermodynamics Bethe Ansatz equations \cite{Patu2016}  and a two-body calculation, it has been shown\cite{Capuzzi2020}  that  it is possible to obtain a lower bound and an upper bound for 
the rescaled grand-canonical contact $\mathcal{C}_N(g,T)^{gc}/N^{5/2}$ or for the canonical one
$\mathcal{C}_N(g,T)^{c}/(N^{5/2}-N^{3/2})$. These two curves are shown in Fig.~\ref{fig-su2-gc-c}.


\section{Conclusions and outlook}
In this review we have illustrated the various techniques to obtain and exploit the exact solutions for strongly-interacting one-dimensional trapped bosons,  fermions and mixtures.
Infinite interactions play the same role as the Pauli principle, allowing the mapping of the many-body wavefunction for the strongly correlated many-body system onto that for a system of spinless non-interacting fermions.
The knowledge of the exact wavefunction gives a unique opportunity to unveil the properties of strongly correlated one-dimensional fluids. 
It also  allows to understand of the role of particle-exchange symmetries in the mixtures. 
Moreover, it allows to benchmark both classical numerical simulators, usually used for systems at finite interaction and temperature, as well as particular experimental setups for quantum simulators. One example is provided by the mapping of strongly interacting fermions onto a spin chain \cite{Deuretzbacher2014,murmann2015antiferromagnetic,Deuretzbacher2017a,Deuretzbacher2017}.
The exact solution allows also to provide tests of other approximate approaches, as the Luttinger liquid solution, as done eg in  Refs.~[\onlinecite{Didier09a}], and [\onlinecite{polo2018damping}].

The detailed study of the predictions of the Girardeau mapping has yielded a wealth of information on the properties of the 1D fluids: for example, the study of the dynamical structure factor shows the effects of the curvature of the dispersion of the collective excitation modes and its broadening due to particle-hole excitations; both effects are not included in the usual Luttinger liquid picture, but require non-linear Luttinger liquid tools \cite{imambekov2012onedimensional}. Another striking prediction of the TG solution is the fact that large-amplitude breathing modes in a harmonic trap are not damped. This has stimulated very general reflections on damping and thermalization in closed quantum systems, and about the fate of the system at very long times \cite{rigol2008thermalization}, which were then followed by the experiment on the quantum Newton's cradle \cite{Kinoshita06}. In multicomponent Fermi gases, it has been demonstrated that the tails of the momentum distribution are fixed by the symmetry of the mixture \cite{Decamp2016-2} thus providing a new type of symmetry spectroscopy. Furthermore, the dynamics of the magnetization of strongly repulsive SU(2) fermions in harmonic trap \cite{pecci2021universal}  points to a connection to the Kardar-Parisi-Zhang universality class \cite{ljubotina2017spin,ljubotina2019kardar}, a statistical physics model describing the growth and roughening of classical interfaces \cite{kardar1986dynamic}.

Several directions open up in this research field. First of all, 
even if we have shown that these techniques can be applied not only for the case of zero-temperature systems at the equilibrium, but also for the case of finite temperature and for the full quantum dynamics, there is
a real challenge to find new strategies in order to be able to deal with the amazing increasing complexity
arising when more and more single-particle orbitals have to be included in the calculations.
In particular, there is a clear need to  improve the existing solution  strategies in order  to reach eg larger system sizes, or describe arbitrary temperatures. In this respect, it is very useful to share open source codes as done eg in [\onlinecite{loft2016conan}] and [\onlinecite{deuretzbacher2016momentum}].
Secondly, the solutions illustrated in this review could be used to explore further the physical properties of  correlated gases. This is extremely useful since these solutions are amongst the very rare cases where one can follow exactly the arbitrary dynamics even at long times.  This will allow to describe specific dynamical protocols useful for quantum information and quantum state engineering, or predict the outcome of novel quench protocols. For example, a quantum heat engine was recently designed exploiting the TG solution \cite{fogarty2020manybody}, and quantum simulation of the spin-Seebeck effect was proposed by exploiting the mapping to the inhomogeneous Heisenberg chain \cite{Barfknecht2021}. 
Finally, the quest is still open to 
find other exact solutions, as eg for the case of particles of unequal masses \cite{Loft2015-var,Dehkharghani2016-unb,scoquart2016exactly,harshman2017integrable}.   
\section*{Data Availability}
The data that support the findings presented in this review are available from the corresponding authors of the original publications upon reasonable request.

\acknowledgements
We are grateful to M. Albert, T. Busch, D.M. Gangardt, J. Polo Gomez, M. Olshanii, G. Pecci, L. Santos and N.T. Zinner for  their comments and suggestions on the manuscript. We acknowledge funding from the ANR-21-CE47-0009 Quantum-SOPHA project.

\section*{Author Declarations}
\subsection*{Conflict of interest} The authors have no conflicts to disclose.


%

\end{document}